\documentclass[sageapa,Afour,times]{sagej}

\usepackage{moreverb,url}

\usepackage[colorlinks,bookmarksopen,bookmarksnumbered,citecolor=red,urlcolor=red]{hyperref}

\newcommand\BibTeX{{\rmfamilyhttps://web.archive.org/web/20191218050603/https://about.fb.com/news/2019/12/helping-fact-checkers/ B\kern-.05em \textsc{i\kern-.025em b}\kern-.08em
T\kern-.1667em\lower.7ex\hbox{E}\kern-.125emX}}

\usepackage{pgfplots}
\pgfplotsset{compat=1.17} 
\usetikzlibrary{pgfplots.groupplots}
\usetikzlibrary{shapes.geometric}
\usepgfplotslibrary{fillbetween}
\selectcolormodel{cmyk}

\usepackage{graphicx}
\usepackage{doi}
\usepackage{placeins}
\usepackage{longtable}

\newcommand{\rev}[1]{#1}

\def\volumeyear{2016}

\begin{document}

\runninghead{Resnick et al.}


\title{Searching For or Reviewing Evidence Improves Crowdworkers' Misinformation Judgments and Reduces Partisan Bias}

\author{ Paul Resnick\affilnum{1}, Aljohara Alfayez\affilnum{1}, Jane Im\affilnum{1}, and Eric Gilbert\affilnum{1}}

\affiliation{\affilnum{1}University of Michigan, 	Ann Arbor, Michigan, US}

\corrauth{ Paul Resnick, University of Michigan, 	Ann Arbor, Michigan, US.}

\email{presnick@umich.edu}


\begin{abstract}
Can crowd workers be trusted to judge whether news-like articles circulating on the Internet are misleading, or does partisanship and inexperience get in the way? 
And can the task be structured in a way that reduces partisanship?
We assembled pools of both liberal and conservative crowd raters and tested three ways of asking them to make judgments about 374 articles. In a no research condition, they were just asked to view the article and then render a judgment. In an individual research condition, they were also asked to search for corroborating evidence and provide a link to the best evidence they found. In a collective research condition, they were not asked to search, but instead to review links collected from workers in the individual research condition.
Both research conditions reduced partisan disagreement in judgments.
The individual research condition was most effective at producing alignment with journalists' assessments. In this condition, the judgments of a panel of sixteen or more crowd workers were better than that of a panel of three expert journalists, as measured by alignment with a held out journalist's ratings. 
\end{abstract}


\maketitle

\section{Introduction}
The rise of misinformation on social platforms is a serious challenge. About a fifth of Americans now obtain their news primarily from social platforms \cite{mitchell2020americans}; however, because anyone can post anything, a user can easily encounter misinformation \cite{vosoughi2018spread,grinberg2019fake}. 
Platforms have recently committed to countering the spread of harmful misinformation, via methods such as informing news consumers that an item may be false or misleading, algorithmically demoting it, or removing it entirely \cite{fb3rdparty}. 

\rev{The ground truth of whether a news article merits one of these enforcement actions is not simple to define. Typical news articles contain multiple factual claims, some of them more prominent than others, some of them more factually accurate than others, and some of them more harmful than others in their impacts on readers who may be misled. Moreover, an article may mislead readers without making a factual claim, such as by asking a factual question in a way that invites an incorrect inference about the correct answer. Thus, assessments of misinformation often go beyond a binary notion of true and false to use multi-point scales and to include aspects such as misleadingness, harm, neutrality, completeness, accuracy, trustworthiness, and objectivity \cite{SOPRANO2021102710,barbera2020crowdsourcing,allen2020scaling}.
}

Crowdsourcing, or using pools of laypeople, has been shown to be effective for many labeling tasks (e.g., \cite{mitra2015credbank,chung2019efficient}). 
\rev{Social media platforms have begun to explore ways to use crowd workers and platform users as part of their misinformation assessment processes. For example, Facebook implemented a process they called ``Community Review'' where crowd workers identified misinformation to route for final evaluation by fact-checking organizations \cite{fbcommunityReview}. Twitter's Birdwatch program created a community of Twitter users who identify and label tweets that propagate misinformation \cite{colemanIntroducingBirdwatchCommunitybased2021}. Any process that involves broad participation, however, raises several concerns.
}

One concern is that lay raters will make judgments based on ``gut reactions'' to surface features, rather than determining the accuracy of factual claims. After all, the reason that misinformation circulates widely on social media platforms is that users upvote and share it, often without considering it deeply or even reading it \cite{pennycook2019lazy}, which is exacerbated by social media encouraging engagement more than accuracy \cite{jahanbakhsh2021}. There is reason, however, to be optimistic that this can be overcome. Priming people to focus on accuracy, people's intention to share other articles became more correlated with the accuracy of those articles \cite{pennycook2019understanding}. Thus, when asked to rate whether an article contains misinformation, people may assess it differently than they do when they choose whether to click or share an item. 

Even when going beyond gut reactions, lay raters may lack the expertise to assess information in the way journalists and professional fact-checkers do. \cite{wineburg2019lateral} found that journalists assessing an article tend to search for external corroborating or contradictory information, including what the authors call ``lateral reading'' about the author or source as well as the contents. College students and even history professors, by contrast, tended to focus more on signals within an article itself, to the detriment of their assessments. Here, too, however, there is reason for optimism. In another study, they found that with two 75-minute training sessions on searching for and interpreting external signals, students' search practices and reasoning processes improved~\cite{mcgrew2019improving}.

A third concern is that lay raters may be ideologically motivated or biased. There are partisan divides in beliefs about some factual claims that have become politicized, such as the size of crowds at President Trump's inauguration or whether wearing face masks significantly reduces the spread of COVID-19. 
\rev{Results of past research on the effects of partisanship on rater assessments of misinformation are mixed. At the level of sources rather than individual articles,}
\cite{pennycook2019fighting} found that Democratic and Republican crowd workers largely agree when distinguishing mainstream news sites from hyper-partisan and fake news sites, 
\rev{but \cite{michael2021relationship} found that political affiliation influenced people’s beliefs about which news sources are ``fake''.  In other studies, partisanship also affected assessments of whether individual articles were misleading \cite{farago2019we,barbera2020crowdsourcing}.}

Two recent studies elicited misinformation judgments from lay raters about specific articles~\cite{allen2020scaling,article:NYUarticlecollection}. Perhaps surprisingly, they came to quite different conclusions. One found that the average of the ratings of a group of lay raters could identify misinformation pretty well, even though the raters saw just the headlines and ledes of articles and not the entire articles~\cite{allen2020scaling}. The other found that crowds performed worse than a journalist~\cite{article:NYUarticlecollection}. Here we report on a larger study, using articles from both of the other studies. 
\rev{Thus, one contribution of this study is to resolve the apparent contradiction in findings, which we address in detail in the discussion section.}

More importantly, we study the effects of varying the elicitation process. 
In a control condition, raters assessed whether articles were false or misleading after opening the articles but without doing any additional research. In one treatment condition, which we call the individual research condition, each rater also searched for corroborating evidence. This was intended to elicit ``informed'' judgments rather than gut reactions. 
\rev{A version of this approach has been used in Facebook's Community Review \cite{fbcommunityReview} and in academic studies \cite{barbera2020crowdsourcing,SOPRANO2021102710}, but no study has examined whether searching for corroborative evidence improved rater performance.}  
In the other treatment condition, \rev{which we call collective research}, raters consumed the results of others' searches. This was intended to reduce ideological polarization---we deliberately included corroborating evidence links discovered by both liberal and conservative raters in an effort to broaden the search horizon for any given rater. Both of the treatment conditions thus enforced some form of lateral reading.

\rev{We assessed performance in three ways. First is inter-rater agreement, a measure of internal consistency. Second was partisan disagreement, measured by the correlation, across articles, between the mean rating of liberal raters and the mean rating of conservative raters for each item.}

\rev{The final performance measure for an MTurk panel was agreement with a journalist's ratings of the same articles. While the assessment of misinformation cannot be reduced to a purely objective binary distinction between truth and falsehood, as argued above, there is good reason to treat journalists' assessments as the best available indicator of an underlying ``ground truth'', what some refer to as an ``alloyed gold standard'' ~\cite{10.1093/oxfordjournals.aje.a009089}.  
Fact checkers have well-codified processes and criteria focusing on specific factual claims. Fact-checkers and journalists more generally have a strong ethos of trying to be unbiased, separating their personal opinions from their professional judgments~\cite{mena2019principles,rodriguez2021debunking}. 
Moreover, judgments on some content may require domain expertise or an accumulated knowledge of the current disinformation actors and the larger narratives they are promoting \cite{10.1145/3415204}, expertise that journalists accumulate by virtue of spending more time than the general public in following the news.}

\begin{figure*}
    \centering
    \includegraphics[width=0.9\linewidth]{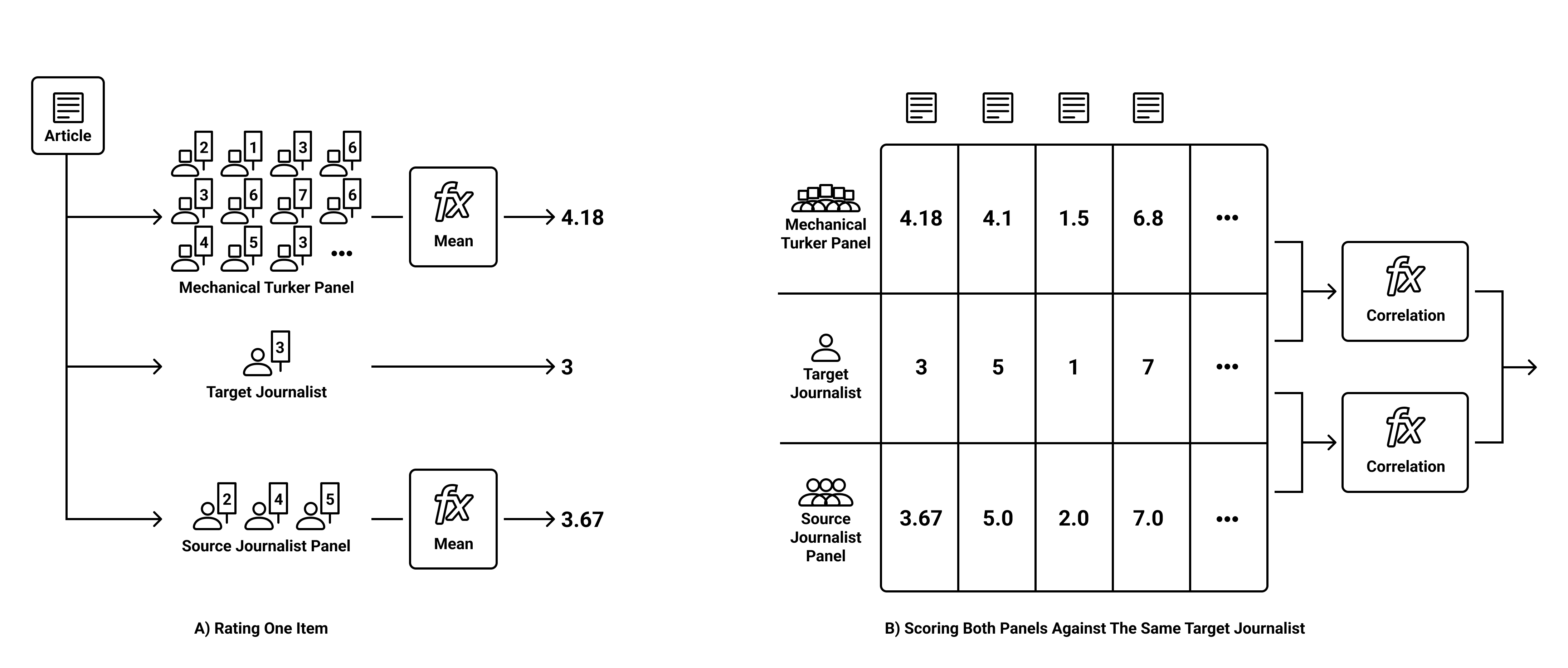}
    \caption{Conceptual diagram of the performance benchmarking process. A) Rating a single item; compute the mean rating of a Mechanical Turk panel and the mean rating of a source journalist panel, and collect a single target journalist's rating. B) Both turker and journalist panels are scored against the same held-out journalist, by computing the correlation over many items.}
    \label{fig:rater-equivalence-flow}
\end{figure*}

\rev{As a comparison point, we also benchmarked the performance of MTurk panels against simulated subsets of one, two, or three journalists. To enable an apples-to-apples comparisons, all rater panels, whether lay raters from Mturk or journalists, were scored by correlating their mean ratings with those of a single held-out journalist, as illustrated in Figure~\ref{fig:rater-equivalence-flow}.}


\rev{Comparison to the performance of simulated panels of journalists is especially interesting because platforms may face the practical question of whether and when to rely on crowd judgments to extend the reach of journalists' or fact checkers' judgments, which are available for only a limited set of items. If a platform would be willing to rely on the judgment of a single journalist, then arguably it should be willing to rely on a crowd that performs better than a single journalist at predicting what another journalist would say. If a platform would be willing to rely on the majority rating of three journalists, then arguably it should be willing to rely on a crowd that performs better than a panel of three journalists at predicting what another journalist would say.}

\section{Experiment Design}

We conducted a between-subjects study, where each lay rater only experienced one condition. Each rater could rate as many articles as they wanted. Participants were recruited and completed the labeling tasks through MTurk. 

\textbf{Condition 1: No research.} The control condition asks participants to visit a URL (by clicking on a link in the interface) and then rate how ``misleading'' it was on a scale from 1=not misleading at all to 7=false or extremely misleading.

\textbf{Condition 2: Individual research.} The second condition adds an additional task before rating: search for supporting or challenging evidence, and paste a link to that evidence into the rating form. Otherwise, it was identical to the first condition.

\textbf{Condition 3: Collective research.}
In the third condition the rater reviews links to corroborating or challenging information that raters from Condition 2 submitted, rather than searching for their own. We posted rating tasks for an article in Condition 3 after all Condition 2 ratings were received. For each article, we selected up to four links that were submitted by Condition 2 raters: the most popular link among liberals, the most popular among moderates, the most popular among conservatives, and the most popular overall. Often the same link was popular in more than one group, leading to fewer than four total URLs.

The study was was approved by the University of Michigan IRB under HUM00171025.

\subsection{The Rating Apparatus}
The second author conducted iterative usability studies on the survey software itself and the qualification tasks over the course of several months, with pilot participants thinking aloud while using the labeling software. Insights into question wording, presentation and order, and interface controls led to many modifications before data collection began.

Figure~\ref{fig:screenshot-C1} in the Appendix shows the labeling interface in the no research condition.
\rev{The rater assessed how misleading the news item was, on a seven-point scale. Other research has explored the use of coarser or finer scales \cite{barbera2020crowdsourcing} and the use of multiple scales to cover distinct aspects such as neutrality, precision, and completeness \cite{SOPRANO2021102710}. In pilot testing, we found that focusing the question on misleadingness rather than truthfulness helped people think about the effect of the article as a whole and that labeling the extreme points as ``not misleading at all'' and ``false or extremely misleading'' was clear enough that workers were able to make judgments most of the time.\footnote{Raters had the option to say they did not have enough information to make a judgment. This option was rarely used (less than $3\%$ in all conditions, see Table~\ref{tab:not-enough-info} in the appendix). Such ratings were excluded when computing averages.} The Facebook Community Review process tried an alternative approach where raters were first asked to describe a single main factual claim in a news item and then assess its factuality \cite{fbcommunityReview}.}

\rev{Each rater also assessed how much harm there would be if people were misinformed about the topic of the news items, again on a seven-point scale. This encouraged raters to report an item as misleading even if it was about a trivial topic. We were concerned that without this separate question, some but not all raters would factor the importance of the topic into their misleadingness assessments, reducing the consistency of ratings.
}

\rev{After completing the assessment questions, each rater was asked to complete questions about their action preferences for the item: whether they thought platforms should remove the item, reduce its distribution, and/or inform readers by adding warning labels. The distribution of these action preferences and their correspondence with raters' assessments is the topic of another paper.}

\rev{Finally, each rater was asked to predict other raters' action preferences. Questions of this type that encourage reflection about what other raters are likely to say have been shown to increase inter-rater consistency ~\cite{shaw2011designing,10.1145/3290605.3300637}. Analysis of whether the prediction question unintentionally influenced raters' reports of their own action preferences and how much information the predictions provide about others' actual preferences is the topic of another paper, also in preparation.}

Figure~\ref{fig:screenshot-C2} in the Appendix shows the additional request made of raters in the ``individual research'' condition. Prior to answering any of the assessment, preference, and prediction questions, they had to search for evidence and paste the search terms used and a link to the best evidence found. 

Figure~\ref{fig:screenshot-C3} in the Appendix shows the interface in the third condition, where participants were asked to click on links found by participants in the second condition and select the one they thought provide the best evidence.

\subsection{News Articles}

Raters labeled two collections of English language news articles, taken from two other studies that were conducted independently in parallel with this one, by teams at other universities \cite{allen2020scaling,article:NYUarticlecollection}. The first collection was selected from among 796 articles provided by Facebook that were flagged by their internal algorithms as potentially benefitting from fact-checking. Among those, the authors of that paper manually selected 207 articles where the headline or lede included a factual claim. That study's focus was how little information needs to be presented to raters in order for them to make good judgments, and participants were presented with only the headline and lede rather than viewing the entire article \cite{allen2020scaling}. Facebook also provided a topical classification of the articles based on an internal aglorithm that they did not share; of the 207 articles, 109 articles were classified by Facebook as ``political''.

The second collection, containing 165 articles, comes from a study that focused on whether raters could judge items soon after they were published~\cite{article:NYUarticlecollection}. It consisted of the most popular article each day from each of five categories\rev{ defined by the authors of that study}: liberal mainstream news; conservative
mainstream news; liberal low-quality news; conservative low-quality news; and low-quality news
sites with no clear political orientation. Five articles per day were selected on 31 days between November 13, 2019 and February 6, 2020.\footnote{In the published report, the authors analyzed results for only 99 of the 165 articles, excluding those from liberal and conservative mainstream sites. They also added an additional 36 articles in a second wave, which was conducted later; those articles were not assessed in our study.} 

Our rating process occurred \rev{later, from March 1 to March 14, 2020.} Four articles from the first collection and two articles from the second collection were removed because the URLs were no longer reachable when our journalists rated them in May and June, leaving a total of 368 articles between the two collections.

\subsection{Crowds of Lay Raters}

\rev{We recruited a large panel of Mturk workers by posting a paid qualification task. The task was available only to workers located in the U.S. Upon starting the qualification task, each} worker  was randomly assigned to one of the three experiment conditions. They used the interface corresponding to their assigned experiment condition to label a sample item, after which they were given two attempts to correctly answer a set of multiple choice questions quizzing them on their understanding of the instructions. They then completed an online consent form, a four-question multiple choice knowledge quiz, and a questionnaire about demographics and ideology. Participants who did not pass the quiz about the instructions, or answer at least two questions correctly on the political knowledge quiz, were excluded. 

\rev{51\% of raters self-reported as female, 48\% self-reported as male, and less than 1\% each non-binary and preferred not to say. By self-reported age, 25\% were 18-29, 37\% 30-39, 21\% 40-49, 11\% 50-59, and 6\% over 60. For self-reported education levels, 9\% had a high school degree or equivalent (e.g., GED), 21\% some college, 11\% an Associate's degree, 42\% a Bachelor's degree, and 17\% a graduate degree. Thus, the participants were somewhat younger and more highly educated than the U.S. population as a whole.}

Table~\ref{tab:recruitment-funnel} shows the recruitment funnel. Because the individual research condition had a more complex task with more complex instructions, more workers dropped out and more failed the instructions quiz.  Thus, in order to get equal numbers of participants who passed the qualification test, more participants had to be randomized to the individual research condition. 
In the discussion section, we return to the question about whether the results may be due to selection of a more qualified or conscientious pool of raters.
Of participants who completed the qualification, a higher percentage in the second condition went on to complete rating tasks.

At the completion of the qualification task, workers were assigned to one of nine groups based on their randomly assigned condition and their ideology. 
We asked about both party affiliation and ideology, each on a seven-point scale, \rev{using two standard questions (VCF0301 and VCF0803) that have been part of the American National Election Studies (ANES) since the 1950s~\cite{anes-documentation}}. Raters who both leaned liberal and leaned toward the Democratic Party were classified as \emph{liberal}; those who both leaned conservative and leaned toward the Republican Party were classified as \emph{conservative}; others were classified as \emph{moderate}.

\rev{The news items were posted at various times over the span of fourteen days, from March 1 to March 14, 2020. For each condition by ideology group, the first eighteen workers to claim the task rated the item. The median time from posting to receiving the last rating was: 17 hours, 29 minutes in the no-research condition; 18 hours, 22 minutes in the individual research condition; 19 hours, 23 minutes in the collective research condition.}

Table~\ref{tab:raters-by-ideology} shows that more raters were liberal than moderate or conservative. \rev{Thus, in order to get eighteen ratings from each rater group, on average each moderate and conservative rater rated more items than each liberal rater, as shown in Table~\ref{tab:ratings-per-rater-by-ideology}. Figure~\ref{fig:item_rated_num} in the Appendix shows the distribution of ratings per rater.}

\begin{table}
\centering
\small
\caption{Recruitment funnel}
\label{tab:recruitment-funnel}
\begin{tabular}{@{}lrrrc@{}}
 & \begin{tabular}[r]{@{}r@{}r@{}}\textbf{C1}\\\textbf{No}\\\textbf{Research}\end{tabular} 
 & \begin{tabular}[r]{@{}r@{}r@{}}\textbf{C2} \\ \textbf{Individual}\\\textbf{Research}\end{tabular} 
 & \begin{tabular}[r]{@{}r@{}r@{}}\textbf{C3} \\ \textbf{Collective}\\\textbf{Research}\end{tabular} 
 &  \\ [1ex]
 \midrule

randomized & 1530 & 2185 & 1583 \\ [1ex]
passed qualification  & 623 & 622 & 622 \\ [1ex] 
rated at least 1 item & 404 & 500 & 397 \\ [1ex]
median ratings \\per pereson & 14 & 14 & 14 \\ [1ex]
mean ratings \\per person & 49.6 & 40.1 & 50.3 \\ [1ex]
max ratings \\by one person & 352 & 368 & 366 \\ [1ex]
\midrule
\end{tabular}
\end{table}

\begin{table}
\centering
\small
\caption{Raters who submitted at least one rating, by condition and ideology}
\label{tab:raters-by-ideology}
\begin{tabular}{@{}lrrrc@{}}
 & \begin{tabular}[r]{@{}r@{}r@{}}\textbf{C1}\\\textbf{No}\\\textbf{Research}\end{tabular} 
 & \begin{tabular}[r]{@{}r@{}r@{}}\textbf{C2} \\ \textbf{Individual}\\\textbf{Research}\end{tabular} 
 & \begin{tabular}[r]{@{}r@{}r@{}}\textbf{C3} \\ \textbf{Collective}\\\textbf{Research}\end{tabular} 
 &  \\ [1ex]
 \midrule
Liberal & 184  & 247  &  177 \\ [1ex]
Moderate & 113  & 107  & 123  \\ [1ex] 
Conservative  & 107 & 146 & 97 \\ [1ex] 
\midrule
\end{tabular}
\label{tab:rater_per_group}
\end{table}

\begin{table}
\centering
\small
\caption{Median ratings per Turker}
\label{tab:ratings-per-rater-by-ideology}
\begin{tabular}{@{}lrrrc@{}}
 & \begin{tabular}[r]{@{}r@{}r@{}}\textbf{C1}\\\textbf{No}\\\textbf{Research}\end{tabular} 
 & \begin{tabular}[r]{@{}r@{}r@{}}\textbf{C2} \\ \textbf{Individual}\\\textbf{Research}\end{tabular} 
 & \begin{tabular}[r]{@{}r@{}r@{}}\textbf{C3} \\ \textbf{Collective}\\\textbf{Research}\end{tabular} 
 &  \\ [1ex]
 \midrule
Liberal & 10  & 11  &  13 \\ [1ex]
Moderate & 20  & 20  & 14  \\ [1ex] 
Conservative  & 15 & 17 & 16 \\ [1ex] 
\midrule
\end{tabular}
\end{table}

\subsection{Journalist Raters}

Four journalists rated all items from both collections. 
All four had just completed a prestigious, selective fellowship for mid-career journalists \rev{and were recruited based on our contacts with the organizers of the program}. Three were U.S.-based and the fourth had covered U.S. politics for many years. One had been through the American Press Institute's fact-checking bootcamp. One journalist did not rate one item. Another journalist did not rate two items and reported that they did not have enough information to make a judgment on 54 others. Such missing ratings were excluded when computing averages and correlations.

The journalists used the labeling interface for Condition 2, which required them to do individual research. \rev{They provided answers to the first two questions, about whether the article was false/misleading and how harmful it would be if people were misinformed on the topic of the article. They were not asked to provide personal opinions about what actions, if any, platforms should take on the article, and were not asked to predict other raters' opinions about actions. Only their answers to the first question, about how misleading the article was, were used in assessing the performance of panels of lay raters.}

\rev{As a robustness check, we also assessed performance against a set of three fact-checker assessments for the 207 articles in the first collection, which were collected by \cite{allen2020scaling}. Their fact-checkers answered Likert-scale questions on seven dimensions, including whether the article was accurate, trustworthy, and unbiased. The seven answers from each fact-checker were averaged to form a composite assessment of the article by that fact-checker. Table~\ref{tab:results-summary-mit-mit-journalists} in the Appendix shows results of this robustness check, which are qualitatively similar to the results to the results for the same items using the journalists recruited for this study. 
}

\section{Measures and Results}

\begin{figure*}
    \centering
    \includegraphics[width=0.9\linewidth]{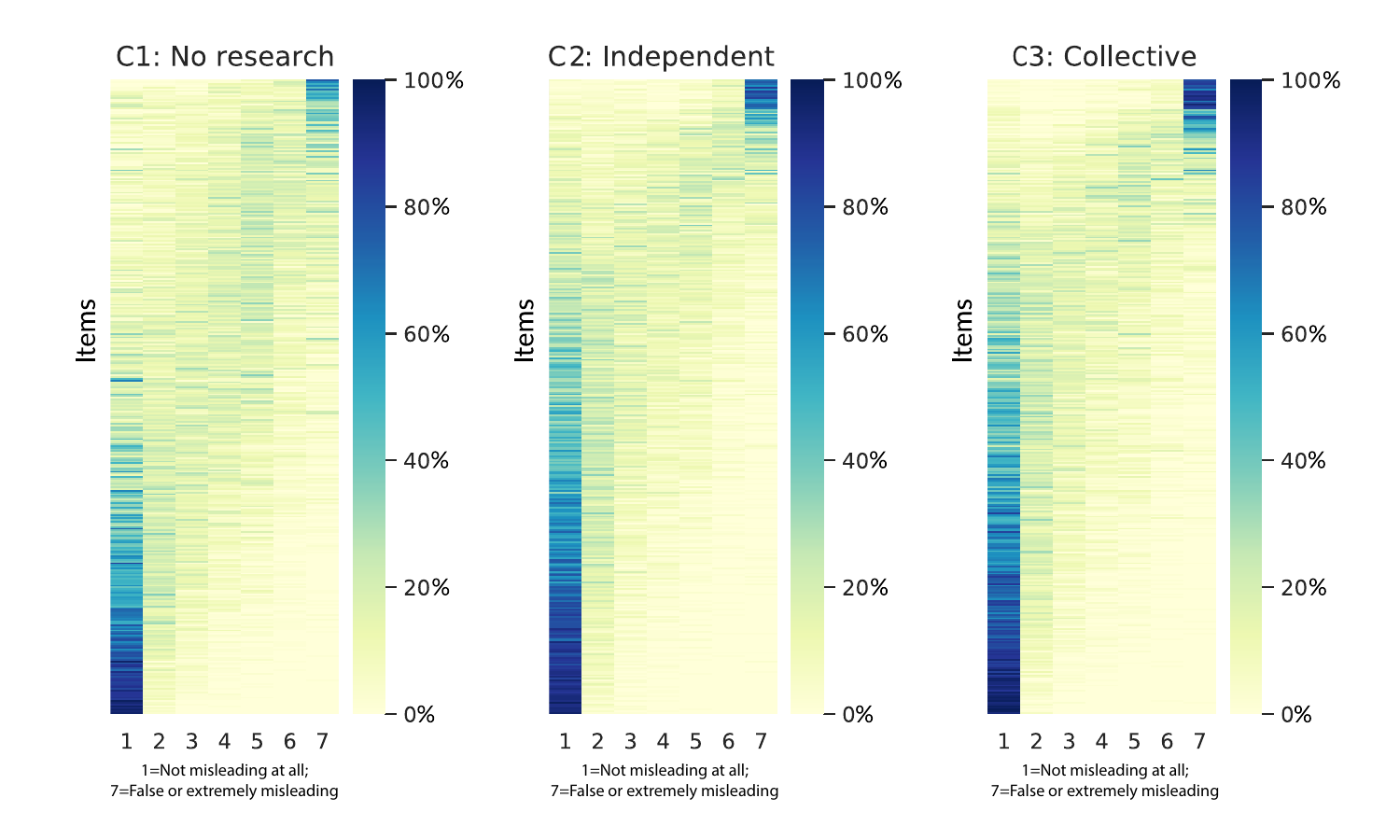}
    \caption{Frequency of the 1--7 ratings for each item in the three conditions. The two research conditions have more items with strong agreement about extreme ratings (1 and 7).}
    \label{fig:heatmap}
\end{figure*}

\begin{table*}[h!]
\centering
\footnotesize
\begin{tabular}{lrrr}
 & \textbf{C1} & \textbf{C2} & \textbf{C3} \\ [1ex]
 \midrule
ICC & 0.422 & 0.458 & 0.510 \\ [1ex]
Liberal-Conservative Correlation & 0.819 & 0.876 & 0.904  \\ [1ex] 
Correlation of one lay rater with 1 journalist  & 0.463  & 0.521 & 0.524 \\ [1ex] 
Lay raters equivalent to 1 journalist & 7.69  & 3.40 & 3.80 \\ [1ex] 
Lay raters equivalent to 2 journalists & $>$54  & 7.08 & 13.55 \\ [1ex] 
Lay raters equivalent to 3 journalists &  $>$54  & 15.22 &  $>$54 \\ [1ex] 
Liberal raters equivalent to 1 journalist & 3.54  & 2.13 & 2.20 \\ [1ex] 
Liberal raters equivalent to 2 journalists & 10.42  & 3.91 & 5.11 \\ [1ex] 
Liberal raters equivalent to 3 journalists &  $>$18  & 6.82 & 13.81 \\ [1ex] 
Conservative raters equivalent to 1 journalist &  $>$18  & 5.48 & 9.68 \\ [1ex] 
Conservative raters equivalent to 2 journalists &  $>$18  &  $>$18 &  $>$18 \\ [1ex] 
Conservative raters equivalent to 3 journalists &  $>$18  &  $>$18 &  $>$18 \\ [1ex] 
\end{tabular}
\vspace{4pt}
\caption{Summary of results by experimental condition.}
\label{tab:results-summary}
\end{table*}

We report three evaluation metrics. First is inter-rater agreement, a measure of internal consistency. Second is ideological polarization, measured by correlation between liberal and conservative ratings. Third is correlation with journalist ratings, which we compare to benchmarks of journalist to journalist correlation.

\subsection{Inter-Rater Agreement}

Figure ~\ref{fig:heatmap} provides a heatmap of the frequency of the 1-7 ratings for each item in the three conditions. Each item is a row and rows are sorted based on the mean ratings for the item across all three conditions. The color coding makes it obvious that in the two research conditions there were many more items with a consensus of 1 (not misleading at all) or 7 (false or extremely misleading), and fewer intermediate ratings.

As a summary measure of agreement, we report the intraclass correlation of ratings ($ICC$)~\cite{shrout1979intraclass}. The $ICC$ is computed with a one-way random effects model, using the mean of raters and consistency as internal metrics \cite{koo2016guideline}. This formulation of the $ICC$ accounts for unstable rater pools: i.e., there is no guarantee that the same group of raters rates each URL. We compute the $ICC$ using the CRAN implementation\footnote{\url{https://cran.r-project.org/web/packages/ICC/ICC.pdf}}. 

\rev{The $ICC$ package also computes 95\% confidence intervals using a bootstrap procedure. For calculating confidence intervals, we specified the ``THD'' method, which is the exact confidence limit equation in \cite{searle1971linear}, because our data is unbalanced \cite{thomas1978interval}. It is unbalanced because we exclude ratings where the rater said they did not have enough information to assess the item, leaving some items with fewer than 54 ratings.}

\rev{The ICC was .422 (CI=[0.388, 0.460]) in the no-research condition, .458 (CI=[0.423, 0.496]) in the individual research condition, and .510 (CI=[0.473, 0.548]) in the collective research condition. The ICC for C1 falls outside the confidence intervals of the two research conditions. The ICC for C2 falls outside the confidence for C3 but just inside the confidence interval for C1.
}

\subsection{Agreement between Partisan Groups}
We also assess whether there is systematic disagreement between raters with different political ideologies. For each item, we compute the mean rating among the eighteen liberal raters and the mean rating among the conservative raters. We then compute the Pearson correlation coefficient, across items, of these mean ratings. 

\rev{We use the percentile bootstrap procedure~\cite{diciccio1988review} to estimate confidence intervals for these correlations. In each of 500 bootstrap samples, we randomly select, with replacement, from the complete set of items, an item set of equal size. For each bootstrap set of items, we follow the procedure above, computing the liberal-conservative correlation coefficient, across items, of the mean ratings. We then compute an interval that covers the middle 95\% of correlation scores that were computed.} 

\rev{The liberal-conservative correlation was .819 (CI=[.782-.851]) in the no-research condition, .876 (CI=[.846-.899]) in the individual research condition and .904 (CI=[.878 -.926]) in the collective research condition. The liberal-conservative correlation in the collective research condition was higher than in the no-research condition for all bootstrap items samples and was higher than in the individual research condition for 99.0\% of the item samples. The correlation was higher in the individual research condition than the no-research condition on 99.8\% of the item samples. 
}


The effects, however, were not uniform for all kinds of items. On just the political items from the first collection, the liberal-conservative correlations in the three conditions were .69, .81, and .79. On just the non-political items from that collection, the correlations were .90, .93, and .98. On the items from the second collection, the liberal-conservative agreement hardly varied between conditions: .83, .83, and .85. See the Appendix for details about results on subsets of the items.

\subsection{Agreement with Journalists}
Our external validity metric compares the performance of panels of lay raters to the performance of panels of journalists. We always score panels, whether MTurk panels or panels of journalists, against a single held-out journalist. \rev{To reduce dependence on any one journalist's ratings, each of the four journalists is used as the held-out journalist and final scores are averaged.}



If, by contrast, we scored panels against the mean of several journalists rather than against a single journalist, the correlation scores would be higher. For example, considering all four possible simulated panels of three journalists, the average correlation with the remaining journalist was 0.73. For simulated panels of two journalists, the average correlation with one of the other two held-out journalists was 0.70
For single journalists, the average correlation with another journalist was 0.65. \rev{It is the relative scores of different panels that is of interest, however. Scoring against a larger target panel of journalists, would raise the correlation scores for \emph{all} panels, both mturk and journalist, but should have little effect on the relative ordering of scores among panels.} 

\rev{The advantage of scoring all panels against just a single journalist is that we leave three of the four journalists available for inclusion in simulated benchmark panels. Thus, we are able to compare the performance of large MTurk panels to the performance of groups of up to three journalists.}

We construct simulated MTurk panels of various sizes, from one to fifty-four, by taking subsets of the 54 MTurk ratings available for each item.  This produces a ``power curve''~\cite{resnick2021survey} as shown in Figure ~\ref{fig:survey-equivalence}.  \rev{For each point on the power curve, the x-value is the number of MTurk ratings to randomly select for each item; the y-value is the expected correlation of the mean of that many ratings for each item with the ratings of a randomly selected journalist (in other words, the power of a panel of that size to predict what a journalist will say). The expected correlation is computed by averaging the results over many runs of the process, using each of the four journalists used as the ratings to predict, and using 200 MTurk subsets of each size (all subsets of a given size if there are less than 200 such subsets).}

\rev{We estimate confidence intervals for the points on the power curve, again using the percentile bootstrap procedure~\cite{diciccio1988review}. In each of 500 bootstrap samples, we randomly select, with replacement, from the complete set of items, an item set of equal size. For each, we compute the power curve points as described above, the expected correlations with a randomly selected journalist for simulated rater panels of different sizes. We then compute a 95\% confidence interval for each point on the power curve as the interval that covers the middle 95\% of expected correlation scores that were computed.}

The intersection points in the graphs in Figure~\ref{fig:survey-equivalence} show the number of lay raters required to get the same predictive power as \rev{benchmark panels of one or three journalists}. For example, in the no-research condition, 7.69 lay raters were sufficient to achieve the same power as one journalist\footnote{Non-integer equivalence values can be interpreted as a randomization procedure. For example, 7.69 lay raters corresponds to using eight raters for $69\%$ of the items, and seven raters on the other items.}.

\begin{figure*}
    \centering
   \pgfplotsset{
    small,
    legend style={
        at={(0.01,0.01)},
        anchor=south west,
    },
   }%
   \pgfdeclareplotmark{fat-}
{%
  \pgfsetlinewidth{0.1pt}
   \pgfsetplotmarksize{.15ex}
  \pgfpathmoveto{\pgfqpoint{\pgfplotmarksize}{0pt}}%
  \pgfpathlineto{\pgfqpoint{-\pgfplotmarksize}{0pt}}%
  \pgfusepathqstroke
}%
\pgfplotsset{
/pgfplots/error bars/error bar style={ultra thin},
/pgfplots/error bars/error mark={fat-}
}
    \begin{tikzpicture}
\sffamily
\begin{axis}[
title = {C1: No research},
title style={align=center,yshift=-.1in},
legend style={font=\scriptsize,
	nodes={scale=1, transform shape},
	at={(0.0,1)},
	anchor=north west,
	draw=none,
	fill=none},
legend cell align={left},
width = 2.0in, height = 2.5in,
ylabel near ticks,
ylabel = {\small Correlation with held-out journalist},
xlabel near ticks,
every tick label/.append style={font=\scriptsize},
xmin=0,xmax=55,ymin=0.4,ymax=0.9,
xtick={0, 7.686232506314485, 27, 54},
tick pos=left,
xlabel={\small Number of raters},
xlabel style = {yshift=0.05in},
yticklabel style={
		/pgf/number format/fixed,
		/pgf/number format/precision=5
},
scaled y ticks=false
]

\addplot[solid, mark=o, mark options={scale=.3}, black]
plot [error bars/.cd, y dir = both, y explicit]
table[y error index=2]{
1	0.46259209556674286	0.040789089651074084	0.03477983984260191
2	0.5403854434381433	0.04298457612574458	0.035796764770371414
3	0.585586724384472	0.04421999384100095	0.03642706614556113
4	0.6116541254139928	0.04430686259226846	0.03562377697799679
5	0.6230495352469727	0.044887002980192436	0.03648722956394845
6	0.6325391980511068	0.04443030362193967	0.03603256211186956
7	0.6428851572055717	0.04462851823913594	0.03645502268272938
8	0.6523705252941828	0.045092347079694006	0.036863777936612774
9	0.6571572820738834	0.04414270557189326	0.0363115205330079
10	0.6625226880569659	0.04442906422802695	0.03616408482079558
11	0.6674310591594415	0.04399074955590454	0.036548706405826636
12	0.667563142272071	0.045368962998895035	0.03644853084060873
13	0.6723791436728708	0.044484916212478076	0.0361345074085736
14	0.67293303643727	0.04521157368677653	0.03632739393329898
15	0.677126781234325	0.045127013611712896	0.036246782489973595
16	0.6770269271499978	0.044932324380055566	0.03652791921057574
17	0.6792426684356743	0.04498646247544191	0.03652865129493632
18	0.6818910661420717	0.04454730723112288	0.03626995499538599
19	0.6833214875585567	0.04477014777338717	0.0366888185922446
20	0.6835827928393782	0.04487746487520328	0.036502136696572096
21	0.6837034747724233	0.044970153672407975	0.0361312847412788
22	0.6857229792993706	0.04456745684606256	0.0363427743821475
23	0.6856550757395634	0.04497397423886729	0.03637039917398488
24	0.6862235860100638	0.04498909986635091	0.03638105717270812
25	0.6885590664240229	0.04486055816558954	0.03617730210580561
26	0.6880480467843314	0.04475641353099713	0.03650923240232762
27	0.6897423512618753	0.044722423905234954	0.03657163374901573
28	0.6893186842044053	0.044855802110073895	0.036445784721258034
29	0.6897081303657424	0.04490609223814379	0.03648897817370089
30	0.6905627064133655	0.04473370275147104	0.036423550771728874
31	0.691305393892613	0.0448768905328506	0.03643896418272663
32	0.6921368263959684	0.044341109309765536	0.0364408158599262
33	0.692645169466738	0.04476663132582592	0.036420284997251584
34	0.6916691172143389	0.044884661151507355	0.036381424196910817
35	0.6935207341071861	0.04479234649225461	0.03634012030533229
36	0.693606278362587	0.044461897599096134	0.036392917828826366
37	0.6939659441675421	0.04493683296186057	0.03607388345955753
38	0.693901495699461	0.04471470757522089	0.036299993594262836
39	0.694339282971447	0.04467636588367074	0.03635134293000086
40	0.6942668900625995	0.044726339679449945	0.03621895479767878
41	0.6953024559112403	0.04470652777808726	0.036203198707767004
42	0.6952762670076733	0.04461854725731462	0.03619053566604835
43	0.6950609359939058	0.0447896614353408	0.03609378349796366
44	0.6954436199229884	0.04469103037947886	0.036232843659818825
45	0.6953589796248562	0.044686081399106015	0.03637148674468027
46	0.696142131084757	0.044610890729582686	0.036248118969308574
47	0.6962120554195057	0.044500160985821324	0.036199192819294845
48	0.6967855237279206	0.04441998084660381	0.036227290714215044
49	0.6964988894072335	0.04459013018511471	0.03633157906111073
50	0.6968465071865485	0.04457219226295561	0.03616418552094902
51	0.6968470663724422	0.04447020672000945	0.03619199511162596
52	0.696897609168747	0.04455275525533642	0.03623513354104968
53	0.6972980962984167	0.044554204973381584	0.03620950105582099
54	0.6974584520841695	0.044540492677371146	0.03620893977092132

};
\addlegendentry{Turkers}

\addplot[mark=none, blue, dashed, thick, samples=2, domain=-1:55] {0.7331056582890765};
\addlegendentry{three journalists}

\draw [blue, fill=blue, opacity=0.1] (axis cs:-1,0.6867234039011271) rectangle (axis cs:55,0.7708079395395433);

\addplot[mark=none, green, dashed, thick, samples=2, domain=-1:55] {0.6493943251223347};
\addlegendentry{one journalist}

\draw [green, fill=green, opacity=0.1] (axis cs:-1,0.5941083989904011) rectangle (axis cs:55,0.6954345399612075);

\addplot[mark=*, mark options={scale=.8}, blue, thick]
    table[]{
        54.0	0.7331056582890765
    };

\addplot[mark=none, blue, dashed, thick, samples=2, domain=-1:31] coordinates {(54.0,0.7331056582890765) (54.0,0)};

\addplot[mark=*, mark options={scale=.8}, green, thick]
    table[]{
        7.686232506314485	0.6493943251223347
    };

\addplot[mark=none, green, dashed, thick, samples=2, domain=-1:31] coordinates {(7.686232506314485,0.6493943251223347) (7.686232506314485,0)};


\end{axis}
\end{tikzpicture}
     \begin{tikzpicture}
\sffamily
\begin{axis}[
title = {C2: Independent},
title style={align=center,yshift=-.1in},
legend style={font=\scriptsize,
	nodes={scale=1, transform shape},
	at={(0.0,1)},
	anchor=north west,
	draw=none,
	fill=none},
legend cell align={left},
width = 2.0in, height = 2.5in,
ylabel near ticks,
xlabel near ticks,
every tick label/.append style={font=\scriptsize},
xmin=0,xmax=55,ymin=0.4,ymax=0.9,
xtick={3.3980444023110286, 15.22363252288848, 27, 54},
tick pos=left,
xlabel={\small Number of raters},
xlabel style = {yshift=0.05in},
yticklabel style={
		/pgf/number format/fixed,
		/pgf/number format/precision=5
},
scaled y ticks=false
]

\addplot[solid, mark=o, mark options={scale=.3}, black]
plot [error bars/.cd, y dir = both, y explicit]
table[y error index=2]{
1	0.5206934687072168	0.04161269553904745	0.03679082188644123
2	0.5975760765007267	0.040470738360853464	0.03546290140153996
3	0.6386464457465129	0.042096889625618195	0.03411741172513516
4	0.665648155388503	0.04062195795190804	0.032727963549386785
5	0.6826732045691309	0.03987314745773429	0.03253010962288427
6	0.6956907293821327	0.03897093051459277	0.03300215861801459
7	0.7025075472196471	0.03831469520731312	0.031756722694437745
8	0.7083536647120339	0.03807340801031156	0.03159824089275176
9	0.7147844004116964	0.03887398808988318	0.030974292053477126
10	0.7193417324882819	0.03875741473668448	0.030884100006524084
11	0.7223206954213508	0.03852977541944347	0.030613705236992583
12	0.7250177854036488	0.03830149226680901	0.030636179518878537
13	0.7290730763796478	0.037317389364049824	0.03022354337204447
14	0.7308430180684437	0.037829115669146596	0.030296086997877003
15	0.7327505346020143	0.038121603666977366	0.03008669931848318
16	0.7343385133202942	0.03795601796362713	0.030342688623188008
17	0.7364709470290188	0.03731278467154775	0.029812259058819568
18	0.7377360859922103	0.037351191693792085	0.030299549909240775
19	0.7383966731097006	0.03766206217571366	0.029928953164886862
20	0.7392828021917817	0.03720203986749293	0.029855743539547785
21	0.7398822580549701	0.037206004492462585	0.029988582293249633
22	0.7412354504570681	0.03701368470124644	0.0297775929441898
23	0.7422402474530767	0.0371541831582084	0.0297874407791463
24	0.7430632581639921	0.03671219413762028	0.030144897470938847
25	0.7431989158262592	0.036838892301174875	0.02988974413320089
26	0.7440009854617514	0.03664191176408915	0.029666090962152114
27	0.7450314954499514	0.03663632478386536	0.029863715161942084
28	0.7451802889895967	0.036450784851753015	0.02976825208590217
29	0.7461575492707131	0.03644615515119565	0.029821817047624988
30	0.74668125847454	0.03641870688966442	0.02957676198228998
31	0.7469778441561682	0.036418555062546654	0.029642661958915406
32	0.7476882778731677	0.03625025492548828	0.029487705088821037
33	0.7482973494634833	0.036377234910984546	0.029432312841939035
34	0.7481370630393367	0.036196649489403154	0.02960870277451244
35	0.7487919847122342	0.036391130937747485	0.02954953568648333
36	0.7491684617153527	0.0363624181781238	0.02950851787440334
37	0.7492736696526083	0.036309097523497535	0.02936935978966393
38	0.7498385205699925	0.03585658898864996	0.029426729543156727
39	0.7499088401401238	0.03601806591642398	0.029339613792684682
40	0.7503016335145267	0.03608624390032911	0.029375827171875812
41	0.7506687767939123	0.03586327129310096	0.029379365808161473
42	0.7506147450707895	0.03614761303923808	0.02932154961636879
43	0.7507461683119765	0.036177834687553645	0.02926961747823109
44	0.7513030486402572	0.03595832394620557	0.029206393692330024
45	0.7512158288898456	0.036036100710271746	0.029290759671156907
46	0.7515823110475344	0.03593177999082364	0.029171509304458887
47	0.7519763550003097	0.03593601664200008	0.029180492245069245
48	0.7519175992890513	0.03600154199834893	0.02929522441225607
49	0.7522102484773266	0.03591637320442265	0.02924788027762104
50	0.752173577080971	0.03598605709512048	0.029197104462056478
51	0.7527235140732991	0.03583382934596213	0.029053617353313688
52	0.752669965873431	0.03586261710928984	0.02917372117146677
53	0.7528261854994683	0.035869824890561786	0.02916800078671211
54	0.7529795241287043	0.035857092103563915	0.029156877912708934

};
\addlegendentry{Turkers}

\addplot[mark=none, blue, dashed, thick, samples=2, domain=-1:55] {0.7331056582890765};
\addlegendentry{three journalists}

\draw [blue, fill=blue, opacity=0.1] (axis cs:-1,0.6867234039011271) rectangle (axis cs:55,0.7708079395395433);

\addplot[mark=none, green, dashed, thick, samples=2, domain=-1:55] {0.6493943251223347};
\addlegendentry{one journalist}

\draw [green, fill=green, opacity=0.1] (axis cs:-1,0.5941083989904011) rectangle (axis cs:55,0.6954345399612075);

\addplot[mark=*, mark options={scale=.8}, blue, thick]
    table[]{
        15.22363252288848	0.7331056582890765
    };

\addplot[mark=none, blue, dashed, thick, samples=2, domain=-1:31] coordinates {(15.22363252288848,0.7331056582890765) (15.22363252288848,0)};

\addplot[mark=*, mark options={scale=.8}, green, thick]
    table[]{
        3.3980444023110286	0.6493943251223347
    };

\addplot[mark=none, green, dashed, thick, samples=2, domain=-1:31] coordinates {(3.3980444023110286,0.6493943251223347) (3.3980444023110286,0)};


\end{axis}
\end{tikzpicture}
    \begin{tikzpicture}
\sffamily
\begin{axis}[
title = {C3: Collective},
title style={align=center,yshift=-.1in},
legend style={font=\scriptsize,
	nodes={scale=1, transform shape},
	at={(0.0,1)},
	anchor=north west,
	draw=none,
	fill=none},
legend cell align={left},
width = 2.0in, height = 2.5in,
ylabel near ticks,
xlabel near ticks,
every tick label/.append style={font=\scriptsize},
xmin=0,xmax=55,ymin=0.4,ymax=0.9,
xtick={3.7977154664886665, 27, 54},
tick pos=left,
xlabel={\small Number of raters},
xlabel style = {yshift=0.05in},
yticklabel style={
		/pgf/number format/fixed,
		/pgf/number format/precision=5
},
scaled y ticks=false
]

\addplot[solid, mark=o, mark options={scale=.3}, black]
plot [error bars/.cd, y dir = both, y explicit, ]
table[y error index=2]{
1	0.5244552365916657	0.05043656338749314	0.042646115880665025
2	0.5924458398580356	0.04878639279894781	0.039846798169239084
3	0.6325422612912204	0.04718487044052955	0.038593372481683264
4	0.6536676682051576	0.0461027299182879	0.037505071052175376
5	0.6677764732403334	0.045056213267508194	0.03648214227142543
6	0.6772308842936952	0.04454752250163374	0.03549811164649286
7	0.6826992447172777	0.04483535552801576	0.03568573846496681
8	0.6901376668032956	0.04401868027921407	0.034847264782998266
9	0.6910876228889521	0.043920572657683676	0.034882491631806345
10	0.6941767300872279	0.043958992164278454	0.03489231347093069
11	0.6992909794958984	0.04363701981737789	0.03437449783530888
12	0.6989633941294531	0.0431840096036169	0.03432556512833507
13	0.7013372922923224	0.04353314826850729	0.033473217862319826
14	0.7043242176651441	0.043296307775952725	0.03409179788473826
15	0.705595933916708	0.043185444102082915	0.03376082310086115
16	0.7058032695435763	0.04347553770934609	0.0338518296917516
17	0.7080428424557397	0.043604238785926674	0.03324089714108058
18	0.7087842402166307	0.04345248987027783	0.03371646156494956
19	0.7094177487785076	0.043364221587203766	0.03337815848886272
20	0.7108314533239026	0.04323769848717196	0.03331966805686348
21	0.7113352556932653	0.0430222315037142	0.03361863162494738
22	0.7119470875564347	0.04323081311418686	0.03315361712613274
23	0.7125716860230926	0.042519441279187054	0.03302806601980102
24	0.7136402281974906	0.04276285478374908	0.033323378154577865
25	0.7141014448122195	0.04312364956899273	0.03313657173889295
26	0.7146552666420254	0.04296181893434026	0.033071584199064974
27	0.7144526915640599	0.04276795560078561	0.033184301845515574
28	0.7146900577905048	0.04299758840159862	0.03327501618129969
29	0.7158560857833614	0.04217290130185347	0.03300414807302121
30	0.7161501890651951	0.04263611399182643	0.03302161278763727
31	0.7161401138836428	0.04222873720788278	0.032979840085267886
32	0.7167585015352604	0.04245680678600139	0.032963195427850756
33	0.7168943367631982	0.042545851453942185	0.03288466463885298
34	0.7172740260085435	0.042644661846046206	0.03286885871893397
35	0.7181322089844145	0.04238519511782013	0.032744823016502944
36	0.7185528917514565	0.04242177835321215	0.03272821830700412
37	0.7180194986468535	0.04232859564714342	0.032757833271647274
38	0.7184117657017545	0.04210539285086767	0.0328575424762535
39	0.7183267604867507	0.0422959966666564	0.0327782381952767
40	0.7187489985372717	0.04230847395505011	0.032721000280130585
41	0.7191662294946819	0.04225697557874997	0.03264271861033596
42	0.7193537128848448	0.0422374085110222	0.032553627985774436
43	0.7191323496150352	0.0421750157502111	0.03281370753636825
44	0.7195051919360845	0.04224275250925191	0.0325238363213014
45	0.7198729450370179	0.04210253998567459	0.032566269735319064
46	0.7199032925937141	0.04226845167048732	0.032702489727275075
47	0.7202491459259158	0.04213527680817175	0.03257816974775729
48	0.7201255545938338	0.04209230044374901	0.03256919285272086
49	0.720024218918412	0.04220336663544	0.032685971695064264
50	0.7204141839331883	0.04216817778312165	0.032514150052640245
51	0.720431665681266	0.04214505417072678	0.03259944289880112
52	0.7207349839081713	0.042113801071015367	0.03259147646486338
53	0.72078391909779	0.04209149100263687	0.03254869176286479
54	0.7209015087699076	0.04207899226134215	0.032537136264571775

};
\addlegendentry{Turkers}

\addplot[mark=none, blue, dashed, thick, samples=2, domain=-1:55] {0.7331056582890765};
\addlegendentry{three journalists}

\draw [blue, fill=blue, opacity=0.1] (axis cs:-1,0.6867234039011271) rectangle (axis cs:55,0.7708079395395433);

\addplot[mark=none, green, dashed, thick, samples=2, domain=-1:55] {0.6493943251223347};
\addlegendentry{one journalist}

\draw [green, fill=green, opacity=0.1] (axis cs:-1,0.5941083989904011) rectangle (axis cs:55,0.6954345399612075);

\addplot[mark=*, mark options={scale=.8}, blue, thick]
    table[]{
        54.0	0.7331056582890765
    };

\addplot[mark=none, blue, dashed, thick, samples=2, domain=-1:31] coordinates {(54.0,0.7331056582890765) (54.0,0)};

\addplot[mark=*, mark options={scale=.8}, green, thick]
    table[]{
        3.7977154664886665	0.6493943251223347
    };

\addplot[mark=none, green, dashed, thick, samples=2, domain=-1:31] coordinates {(3.7977154664886665,0.6493943251223347) (3.7977154664886665,0)};


\end{axis}
\end{tikzpicture}
    \caption{Power curves for the three conditions. The x-axis is the number of turkers. The y-axis is the correlation of the mean of $k$ turkers' ratings with a journalist's rating. The green horizontal lines show the correlation of a randomly selected journalist's rating for each item with a held-out journalist (0.65). The blue lines shows the correlation of the mean of three journalists with a held-out journalist (0.73). Only in C2, the independent research condition, did simuated panels of turkers exceed the correlation score of a panel of three journalists.}
    \label{fig:survey-equivalence}
\end{figure*}

Comparing across the three conditions, we can see that lay raters in the two research conditions correlate with a journalist better than do raters in the no research condition and that the individual research condition has greater power than the collective research condition for large groups of lay raters. In the individual research condition, C2, 15.22 lay raters were equivalent to three journalists; even 54 raters were not sufficient to achieve the same power as three journalists in the other two conditions, and in the no research condition, C1, 54 raters were also insufficient to achieve the same power as two journalists.

To assess reliability we considered the results separately for each bootstrap sample of items (see Table~\ref{tab:reliability-of-comparisons} in the Appendix).
For all group sizes, the power of a group of lay raters in the individual research condition was higher than one in the no research condition in all of the 500 bootstrap samples. With just a single rater, the collective research condition had higher power than the individual research condition on $71.5\%$ of the item samples. With groups of five raters, however, the order flipped: individual research condition performed better than the collective research condition on $98.4\%$ of item samples.

In condition C2, with individual research, eighteen lay raters outperformed two journalists on all of the bootstrap item samples and outperformed three journalists on 65.7\% of samples (see Table~\ref{tab:reliability-of-equivalences} in the Appendix). 54 lay raters outperformed three journalists on 97.4\% of bootstrap item samples.

\section{Discussion}

\subsection{How to Elicit Better Judgments}

As noted in the introduction, prior work found that with two 75-minute training sessions, students adopted practices of external search for supporting and challenging information, and lateral reading about the author and source~\cite{wineburg2019lateral}. In our study, none of the raters received explicit training. 
Both of the treatment conditions, however, required raters to seek out or consider external evidence.
This led to improved misinformation judgments, as measured in a variety of ways: judgments in the two research conditions were more internally consistent between raters, showed less partisan divide, and were better correlated with expert journalist judgments. 

The results are mixed, however, about what is the best way to ask raters to consider external evidence. Condition 2, where each person searched individually, had lower consistency among raters and more partisanship than Condition 3, where raters examined evidence links that were provided to them. 
However, when averaging ratings from several raters, the correlation with a journalist was higher in Condition 2, with the difference becoming more and more reliable when averaging across larger rater panels. Each person doing their own search seems to yield noisier individual assessments. Wisdom of crowds models posit that when averaging several judgments it is better for those judgments to be independent~\cite{surowiecki2005wisdom}. Our results are consistent with that; in Condition 3, examining a common set of links to potential corroborating or challenging evidence may have yielded correlated errors in judgment. 
We suspect that the best procedure for eliciting judgments from raters will be some hybrid that encourages both individual search and considering links that have been discovered by others, especially links discovered by people with different ideologies.

\subsection{A Model of Crowdsourced Judgments}
Consider a stylized hierarchical model of individual raters' judgments when those raters are drawn from some defined pool (e.g., journalists, or liberal workers on Mechanical Turk):
\[Judgment \sim Truth + \textit{GroupOffset} + \textit{RaterNoise} \]
\emph{Truth} for an article is an unobserved hypothetical ``correct label'' for an item. The \emph{GroupOffset} for an article is the difference between the Truth and the (also unobserved) mean rating in the rater population.
\emph{Noise} is the deviations of individual raters from that rater population's mean for an item, due to inter-individual differences and factors like fatigue and distractions.

In this model, there is a Journalist Consensus for each item, consisting of the Truth plus possibly a GroupOffset. And then each journalist sees a draw from a distribution centered on the Journalist Consensus for that item. Similarly, in each condition, for each item there is a Liberal Lay Rater Consensus, a Conservative Lay Rater Consensus, and an overall Lay Rater Consensus.

The GroupOffset can be further decomposed into two components. One is due to \emph{imperfect expertise} of the rater pool and the other due to \emph{ideological bias}.

While none of the underlying components, Truth, GroupOffset, and RaterNoise, can be directly observed, this model provides a framework for interpreting many of the results of our study. Most importantly, the relatively high absolute correlations among judgments from random pairs of raters in all conditions suggest that the Truth component of judgments was fairly large; perhaps the notion of truth is not completely broken beyond repair. 

The noise in individual ratings was much higher for lay raters than for journalists. This can be inferred from the correlations between random pairs of raters from each group. The average correlation between a pair of random lay raters in the individual research condition was 0.47. 
For a random pair of journalists, it was 0.65. 

The intuition behind the so-called wisdom of crowds is that, following the central limit theorem, the mean of a large sample of independent draws from a distribution has lower variance than the mean of a single draw~\cite{surowiecki2005wisdom}. In our model, taking the mean of many raters should reduce the RaterNoise. We find just that: the correlation between the means of random collections of eighteen raters in the individual research condition was 0.94, up from 0.47 for pairs of individual raters.

In our study, the reduction in Noise was sufficient to overcome the presumed smaller GroupOffset for journalists because of their expertise and professionalism. To see this, note that the correlation of a journalist with the mean of a group of eighteen lay raters in the individual research condition (C2) was .74, much better than the 0.65 correlation between a journalist and another journalist.

Liberal and conservative lay raters had systematically different GroupOffsets. This follows from the fact that groups of liberal and conservative lay raters correlated with each other less than groups of random lay raters. Even in the independent research condition (C2), where this effect was reduced, the correlation between the means of eighteen liberal and eighteen conservative lay raters was 0.88, short of the 0.94 correlation between the means of random collections of eighteen lay raters. This could reflect ideological bias in judgments by one or both groups, or differences in expertise between the groups. We cannot distinguish between ideological bias and expertise differences, nor can we determine which group's Consensus tended to be closer to the hypothetical Truth.

Consequently, we also cannot determine definitively whether the journalists had any bias in their judgments. Liberal lay raters correlated better with journalists than conservative lay raters did, meaning that the Journalist Consensus tended to be closer to the Liberal Lay Rater Consensus than it was to the Conservative Lay Rater Consensus. Thus, to the extent that liberal lay raters had ideological bias contributing to their GroupOffsets, the GroupOffsets for journalists must reflect similar ideological bias. On the other hand, if the GroupOffsets for liberals reflected only limited expertise rather than ideological bias, then a better correlation with journalists might not imply any ideological bias on the part of the journalists.

\rev{Changes in the design of the rating task can reduce both RaterNoise and the GroupOffset. We interpret the higher inter-rater agreement (ICC) in the two research conditions as an indication that either seeking corroborating information (C2) or considering corroborating information (C3) reduces RaterNoise. The higher agreement in the two research conditions between liberals and conservatives and between both groups and journalists in the two research conditions indicates that seeking or consuming corroborating information brings the GroupOffsets closer to each other; one plausible interpretation is that the research conditions reduced the GroupOffsets for both liberals and conservatives, yielding ratings closer to the underlying Truth.}

\rev{To the extent that GroupOffsets reflect underlying differences of ideology, values, or lived experiences, the design goal may not always be to eliminate GroupOffsets. Instead, it may be desirable to design elicitation processes that simply illuminate systematic differences in assessments. In other content moderation settings besides misinformation, researchers have begun to explore approaches that explicitly try to elicit or predict differences in the ratings of people from different demographic groups~\cite{goyal2022your,gordon2022jury}. 
To the extent that society comes to view judgments about what constitutes harmful misinformation as partly subjective rather than purely objective, similar approaches may be useful.}

\subsection{Use Cases for Crowdsourced Misinformation Assessments}

\rev{How might assessments from lay raters be incorporated into misinformation enforcement practices?}
\rev{Platforms articulate policies and enforcement actions~\cite{MetaMisinformationStandards,TwitterAddressMisinfo,krishnan2021research}.
Though the policies vary and their details are often not made public, at a high level they generally prescribe some kind of enforcement action against posts that are harmfully misleading.
One common enforcement action is to alert users that a particular piece of content may contain misinformation, through a label, color coding, or text. Twitter refers to this as warning~\cite{twitter-warning-def}. Facebook refers to it as an inform action~\cite{lyonsHardQuestionsWhat2018}. Another possible action is to downrank a content item so that it appears later in search results or news feeds and thus fewer people encounter it. Many other such strategies are also available in addition to filtering or removing the content~\cite{LoModerationInventory}.}

\rev{Fact-checks of specific claims, made by external organizations (third-party fact-checkers, as Facebook describes them), serve as an important resource in determining whether particular claims are harmfully misleading~\cite{fb3rdparty}. The external organizations do not, however, make judgments about whether individual news articles or social media posts are promoting any particular claims. The external organizations also investigate only some of the claims that circulate. For moderating particular articles and posts, platforms rely on a combination of user flagging and automated algorithms to surface some content as potentially problematic. Sometimes moderation actions are taken automatically; in other cases, paid human moderators review the posts. Current practices at platforms constrain the human moderators to apply very detailed policy guidelines that identify particular classes of claims as constituting harmful misinformation. For example, Twitter's COVID-19 misinformation guidelines specified a policy against tweets that ``misrepresent the protective effect of vaccines"~\cite{twitter-COVID-19-policy}.}

\rev{One potential role, then, for lay assessments of misinformation is in that final step, where human moderators review individual posts.  Moderators could be given more leeway to identify any harmful misinformation, even on topics which have not yet been codified into platform-specific taboo claims and even for content that external organizations have not (yet) published fact-checks for. 
For this use case, it is important to understand both the quality of ratings that are produced and the costs. The results of this study, together with that of ~\cite{allen2020scaling}, demonstrate a quality-cost tradeoff. As raters move from the low-cost activity of just looking at headlines to the higher cost of scanning the entire news article to the highest cost activity of searching for corroborating evidence, they produce higher quality ratings, with more inter-ideology agreement and more agreement with expert raters. 
}

We offered raters in the no research condition \$.50 for each article rated and in the two research conditions \$1 for each article rated, in order to compensate them for the extra time required to do research. If, in practice, one's goal was to get a rating of quality similar to what one would get from a single journalist, this could have been accomplished for \$5.50 using eleven raters in the no research condition or \$4 using four raters in either of the research conditions. 
\rev{We note that costs in this range are probably not sustainable if even a tiny fraction of all posts are sent to human moderators. If, on the other hand, only news URLs were evaluated and not every post that linked to one, this process could potentially scale to that level of use.}

\rev{Facebook's Community Review project described using paid crowd workers in a different way. In that project, their role was to prioritize content for routing to the external, third-party fact checkers~\cite{fbcommunityReview}. For that use case, it is probably not necessary to approach or exceed the performance quality of a journalist, since content flagged as problematic will subsequently be reviewed by external fact-checking organizations before any action is taken. Thus, lower cost processes such as only examining headlines, and using smaller numbers of raters, may be sufficient. If, however, the crowd judgments would be used to determine enforcement actions in the interim, before external fact-checking organizations rendered final judgments, then a higher cost process that required raters to search for corroborating information would be more appropriate.}

An alternate use-case, which we favor, conceives of using citizen juries as a governance mechanism, rather than as a way to speed up decision making~\cite{zittrainJuries,FanZhangJuries}. In that vision, crowd-based misinformation judgments could be used as part of appeals processes, as ground truth for transparency reports about platform performance, and as training data for human and automated processes. For that, they would need to operate on a medium rather than large scale, and the higher cost of requiring raters to search for corroborating evidence, or even engage in extensive deliberation with each other, would be justified. 
\rev{For this use-case, it might be particularly interesting to explore the role that platform users themselves might play in populating the juries. The Twitter BirdWatch system offers an interesting first attempt~\cite{colemanIntroducingBirdwatchCommunitybased2021}, but may suffer from allowing people to self-select which content to review, making it prone to strategic manipulation. SlashDot's system of meta-moderation, where users were assigned things to assess, rather than choosing them, may help to counter manipulation~\cite{lampe2004slash}. 
}

\subsection{Limitations}

One limitation of our study is that Turkers passed the qualification process at a lower rate in Condition 2. Thus, it is possible that the better performance in that condition was due, in part, to a pool of raters who were more diligent or skilled, rather than the requirement that they search for a corroborating source. It would be interesting in a future study to tease apart the selection effect from the task effect; if the selection effect is sufficient to yield the rater performance found in the second condition, without requiring raters to actually perform independent research on each item, the costs of rater labeling could be further reduced. 

Another limitation is that articles were rated weeks to months after they were first posted. It is possible that searching for corroborating evidence would not be as impactful soon after the articles were posted, and thus not as effective at driving raters to provide better misinformation judgements. The study that generated our second collection of articles~\cite{article:NYUarticlecollection} was explicitly designed to compare judgments made by lay raters and journalists within a few days of an article's publication. It would be interesting to analyze how well those ratings correlate with our journalist and lay ratings that were collected several months later.

More training for journalists, more time spent evaluating each article, or incentives for agreeing with each other could lead to higher inter-rater agreement among journalists. That would set a higher benchmark for lay panels to compete against, as noted in~\cite{article:NYUarticlecollection}.

\rev{Finally, we should not assume that asking lay raters to search for corroborating evidence will continue to be as effective in the future as it was in this study. If platforms begin to rely on crowd ratings, following any of the use cases described above, disinformation actors might devise new strategies. Much as they already enact different personas to strategically manipulate left-leaning and right-leaning recipients~\cite{10.1145/3274289}, they may might start strategically planting ``corroborating evidence'' URLs and linking to them in order to boost them in search results.  
}

\subsection{Reconciling Results with Previous Studies}

It is worth revisiting the two previous studies~\cite{article:NYUarticlecollection,allen2020scaling} to try to assess possible reasons for discrepant results. Because we reused most of the same items, we can rule out some possible explanations, while others will require further research to tease apart.

First, we note that the current study and the other two the three studies constructed the journalist benchmark slightly differently. All three studies compared a simulated panel of lay raters to a benchmark simulated panel of one or more journalists. We ensured apples-to-apples comparisons by scoring both lay panels and journalist panels ability to predict a single held-out journalist's ratings.
By contrast, \cite{article:NYUarticlecollection} scored both lay panels and a single journalist against the modal answer of a panel of journalists. However, for scoring the single journalist, they had to exclude that journalist, and thus they scored the journalist against the majority vote of a smaller panel. Most likely, the majority vote of a smaller panel will have higher variance, which would make their single journalist benchmark score be lower than it would be in a fair comparison. However, because of the tie-breaking methods they used (a 2-2 vote of four journalists was treated as ``not misinformation''), it is possible that comparing to a smaller panel may have produced a score that was higher than it would be in a fair comparison.

The main analysis in~\cite{allen2020scaling} had an even larger difference in the metrics for comparing the performance of lay rater panels and the performance of a benchmark single journalist. Lay rater panels were scored against the mean of three fact-checkers. The single fact-checker was scored against a single fact-checker. With that comparison, they found that the lay rater panels scored better than a single fact-checker.
However, in an appendix (Figure S9 in~\cite{allen2020scaling}) they provide data for lay panels scored against single fact-checkers. As expected, the correlation is lower. \rev{In an apples-to-apples comparison, both scored against a single fact-checker, the lay panel had a lower correlation with a single fact-checker than a single fact-checker did.}
Thus, there may be less discrepancy between the results of the two prior studies than first meets the eye.

Another possible explanation for discrepant results is uncertainty due to sampling error. ~\cite{article:NYUarticlecollection} reported on one binary outcome (whether the panel's prediction matched the majority vote of the journalists) for each of 135 articles. The accuracy reported for the majority vote of a random crowd of 25 people was 62\%, and for the single journalist benchmark it was 69\%. But this difference in performance could easily occur by chance even if there was no difference between the two. Suppose, for example, that panels of 25 people and the benchmark journalist both had 65\% prediction accuracy. The 95\% confidence interval for the sampling distribution of 135 draws from a Bernoulli distribution with p=.65 is $\pm .08$.

In our study, we have a larger pool of items. We quantify the uncertainty of our results through bootstrap sampling of 500 simulated article sets. In our individual research condition, panels of seven or more lay raters outperformed a benchmark of a single journalist on all 500 simulated article sets and the full panel of 54 lay raters outperformed a benchmark of three journalists on 98\% of the simulated article sets.

An even bigger source of uncertainty is the individual journalists. We employed just four journalists and the other studies employed three and six. Even one or two who were outliers from their peers could have reduced the average ability of benchmark journalist panels to predict what other held-out journalists would say. Since we included the same items from~\cite{allen2020scaling} in our study, both teams were able to perform robustness checks using the other team's journalist ratings. Tables~\ref{tab:results-summary-mit} and ~\ref{tab:results-summary-mit-mit-journalists} in the Appendix show qualitatively similar results in our study whether we use ratings from our four journalists or from their three journalists for the analysis. Because the journalist ratings from the other study are not available, we have not been able to perform a similar robustness check on the items from the second article collection.

A third possible reason for discrepant results could be differences between the types of items in the two sets. The first article collection was selected to include only articles that included a factual claim in the headline. These might be easier for lay raters to judge. To assess this, we ran our analyses separately on the two article collections. Results were qualitatively similar for both. On the 98 articles from the second collection that were from fringe rather than mainstream sources, the subset which were analyzed in~\cite{article:NYUarticlecollection}, lay raters correlated with a journalist less well but journalists also correlated with each other less well; nineteen or more lay raters outperformed a benchmark of three journalists (see Tables~\ref{tab:results-summary-nyu} and \ref{tab:results-summary-nyu-fringe-items} in the Appendix). We did, however, find that lay rater performance may have been worse on the subset of 109 political items from the first collection, with panels of 15.26 raters matching the performance of two journalists but even the full panel of 54 lay raters failing to match the performance of three journalists, (see Table~\ref{tab:results-summary-political-items} and Figure~\ref{fig:survey-equivalence-political-items} in the Appendix). Because of the smaller sample size, however, confidence intervals are wider, so there is uncertainty about whether performance on these kinds of items is truly different.

\cite{article:NYUarticlecollection} suggest that the timing of when rating happens may be critical to the capabilities of lay crowds. In particular, they collected ratings within three days after articles first appeared. For some news items, follow-up articles or fact-checks might appear later that help to disseminate correct information. Prior to that, lay raters may be poor judges of information quality. Our study and~\cite{allen2020scaling} had both lay raters and journalists assess articles weeks to months after they first appeared. If lay crowds are especially poor at assessing articles soon after they are posted, relative to a benchmark of journalists, that would limit their use for real-time decision-making. As mentioned previously, if lay panels are conceived of as citizen juries, as part of platform governance and transparency procedures, the ability to make real-time judgments may not be as important.

Finally, we note significant differences in the rating procedures between the studies. Our
 finding that more informed raters make better judgments may explain some of the differences in results. The study from which we drew the first collection of articles asked raters to examine only the headline and lede, and in one condition the source, without clicking through to see the whole article \cite{allen2020scaling}. As noted, they found that the average of fifty lay raters had slightly lower performance than one fact-checker in an apples-to-apples comparison. Our first condition provided a little more information to raters, by asking them to examine the full article, and panels of eight or more lay raters had a higher performance than a single journalist. Our second and third conditions, involving individual research or reviewing the results of collective research, provided our raters with even more information, and led to still better judgments. 

It is also worth noting that refinement of the selection process and the user interface and instructions may make a big difference for the performance of lay crowds. Studies of crowdsourced labeling in other domains have found that quality control measures and small amounts of training are helpful~\cite{10.1145/2702123.2702553}. 
For this study, we developed a custom web interface embedded as an iframe within MTurk pages, and went through multiple rounds of UX testing over several months. We also excluded raters who did not pass a simple test of whether they understood the interface and instructions and those who did not correctly answer two out of four knowledge questions. Finally, many MTurk workers are very conscientious, especially if they fear work rejections that will harm their ability to earn money in the future. In this study, in addition to the misinformation judgments, workers were asked to provide a subjective opinion about what enforcement action they thought platforms should take and to make a prediction about other raters' subjective opinions. They were told that their judgments and opinions would not be evaluated, but that if their predictions were too far off they would be disqualified, and this may have encouraged workers to be conscientious. All of these factors may have led the lay crowds in our study to perform better than they might have otherwise.

\section{Conclusion}

Overall, the results show that juries composed of lay raters could be a valuable resource in assessing misinformation. \rev{It is possible for panels of lay raters to exceed the performance of a panel of three independent journalist ratings, as measured by agreement with a held-out journalists' evaluations.}  

\rev{Requiring raters to become more informed before rendering judgments about misinformation reduces the partisanship and improves the quality of their ratings. When raters examined only the headline and lede of an article, a lay panel approached but did not match the performance of a single journalist~\cite{allen2020scaling}. When raters examined the entire article (our condition C1), a lay panel could exceed the performance of a single journalist, but not a panel of two journalists. When raters examined evidence links (our condition C3), a lay panel could exceed the performance of a two-journalist panel, but not a three-journalist panel. And when raters searched for their own evidence links (our condition C2), a lay panel could exceed the performance a three-journalist panel.}

Much work, remains, however, to refine the processes by which rater pools are selected and the research tasks they are asked to perform. Would even more research, or more research done in a different way, further reduce ideological disagreement between liberals and conservatives and further increase alignment with journalist ratings? Moreover, the present work treats all disagreements in judgments as equally important. However, in reality, we know that certain disagreements (e.g., election fraud, the dangers of COVID-19, etc.) have outsized capability to harm society. More work should be done to systematically understand these factors and how to address them.

\begin{acks}
We thank the four journalists who evaluated the articles. We thank the two other study teams whose article collections we used in this paper. In addition to sharing the article URLs, they provided useful feedback at a workshop where study designs were shared and at a second where preliminary results were shared. In particular, we thank Kevin Aslett, Adam Berinsky, William Godel, Amber Heffernan, Quentin Hsu, Jenna Koopman, Nathan Persily, David Rand, Zeve Sanderson, Luis Sarmenta, Henry Silverman, and Josh Tucker. We thank Kelly Garrett and Eshwar Chandrasekharan for early conversations that influenced the design of the study. We also thank Vitaliy Lyapota for ably implementing the custom labeling interface.
\end{acks}

\begin{funding}
Facebook provided funding to the University of Michigan for this study. Facebook also sponsored the two workshops where study teams shared designs and preliminary findings.
\end{funding}

\begin{dci}
One of the authors of this study (Paul Resnick) served as a paid consultant to Facebook from May 2018 to May 2020.
\end{dci}

\bibliographystyle{mslapa}
\bibliography{references.bib}

\clearpage

\section{Appendix}

\subsection{Reliability of Comparisons Between Conditions}

\rev{We use the percentile bootstrap procedure to estimate confidence intervals for the points on the power curve~\cite{diciccio1988review}. In each of 500 bootstrap samples, we randomly select, with replacement, from the complete set of items, an item set of equal size. For each sample, we compute the power curve points, for each number k of turkers, the expected correlations with a randomly selected journalist. We then compute a 95\% confidence interval for each point on the power curve as the interval that covers the middle 95\% of expected correlation scores that were computed.}
As a measure of the reliability of comparisons between conditions, we report the fraction of those samples where the correlation was higher between one condition than another, as shown in Table~\ref{tab:reliability-of-comparisons}.

\subsection{Reliability of Equivalence Values}

For each of our 500 bootstrap samples of items, we computed the power curve giving, for each number k of turkers, the expected correlation with a journalist's ratings.
\rev{Similarly, for each of the bootstrap samples, we also computed the expected correlation of a single journalist with a held-out journalist's ratings, the expected correlation of a two-journalist panel with a held-out journalist's ratings, and the expected correlation of a three-journalist panel with a held-out journalist's ratings.}
\rev{From these, for each bootstrap sample the journalist-equivalence values were computed: how many turkers are equivalent to one, two, or three journalists. }

In Table~\ref{tab:seq-cis}, we report a range that includes 95\% of the values computed for the bootstrap samples. In Table~\ref{tab:reliability-of-equivalences}, we report the fraction of bootstrap samples where k lay raters outperformed m journalists in predicting the ratings of a held-out journalist.

\subsection{Results for Item and Rater Subsets}

As described in the main text, there were 207 news items in the first sample, of which 109 were marked as political. There were 165 items in the second sample.

We have ratings from four journalists for both samples. In addition, the team that assembled the first sample of items collected ratings from three other journalists, as described in~\cite{allen2020scaling}. The journalists did not answer exactly the same question we posed, but \rev{a journalists' answers on several separate questions were averaged to yield a single number} on a 1-7 scale. We were able to treat these as comparable to our journalist ratings after reverse coding them.

In Tables~\ref{tab:results-summary-mit} and \ref{tab:results-summary-nyu} we report key result metrics for the two article collections separately. Table~\ref{tab:results-summary-mit-mit-journalists} reports key result metrics for the three journalists from~\cite{allen2020scaling}. Table~\ref{tab:results-summary-political-items} reports key result metrics for the only the 109 political items.

\begin{figure*}
    \centering
   \pgfplotsset{
    small,
    legend style={
        at={(0.01,0.01)},
        anchor=south west,
    },
   }%
   \pgfdeclareplotmark{fat-}
{%
  \pgfsetlinewidth{0.1pt}
   \pgfsetplotmarksize{.15ex}
  \pgfpathmoveto{\pgfqpoint{\pgfplotmarksize}{0pt}}%
  \pgfpathlineto{\pgfqpoint{-\pgfplotmarksize}{0pt}}%
  \pgfusepathqstroke
}%
\pgfplotsset{
/pgfplots/error bars/error bar style={ultra thin},
/pgfplots/error bars/error mark={fat-}
}
    \begin{tikzpicture}
\sffamily
\begin{axis}[
title = {C1: No research},
title style={align=center,yshift=-.1in},
legend style={font=\scriptsize,
	nodes={scale=1, transform shape},
	at={(0.0,1)},
	anchor=north west,
	draw=none,
	fill=none},
legend cell align={left},
width = 2.0in, height = 2.5in,
ylabel near ticks,
ylabel = {\small Correlation with held-out journalist},
xlabel near ticks,
every tick label/.append style={font=\scriptsize},
xmin=0,xmax=55,ymin=0.4,ymax=0.9,
xtick={0, 5.766977143786355, 14, 21, 28, 35, 42, 54.0},
tick pos=left,
xlabel={\small Number of raters},
xlabel style = {yshift=0.05in},
yticklabel style={
		/pgf/number format/fixed,
		/pgf/number format/precision=5
},
scaled y ticks=false
]

\addplot[solid, mark=o, mark options={scale=.3}, black]
plot [error bars/.cd, y dir = both, y explicit]
table[y error index=2]{
1	0.41688744988017007	0.07486827051991829	0.06695327875041068
2	0.4991036893395927	0.08192988034864956	0.07147103636307889
3	0.553634019253058	0.08170266029153528	0.06668617050127434
4	0.5831312380696051	0.08107422750170012	0.0667502840673112
5	0.5935124162053327	0.08117357146120174	0.06769008572028379
6	0.6063520591300283	0.08184969172955547	0.06684004881938366
7	0.620485313834666	0.08179176997541926	0.06564654345632126
8	0.6348475556068817	0.07922898559121849	0.06466954173967465
9	0.6382759017565105	0.07997847785332979	0.0637660776531267
10	0.6473668198461456	0.08073574359738322	0.0643531153239909
11	0.6559807427539706	0.07745232821801395	0.06290213766678188
12	0.6547995662210296	0.07773366311871599	0.06383847757624217
13	0.6598964875383209	0.07804641015680513	0.06418125594857649
14	0.6598864662116655	0.07922205098472357	0.06311520260250236
15	0.6670341427646495	0.07775277475965492	0.06317346350391317
16	0.6657041311469843	0.07850618643190554	0.06304065250731739
17	0.6703563880724461	0.07759898810526433	0.06316003075016718
18	0.6739429263993522	0.0776708138869382	0.06259977507050363
19	0.6760081574749307	0.07768719995058981	0.06197997402486066
20	0.6763353388147323	0.07663791288850652	0.06241002363931991
21	0.6749902682991524	0.07778872825163186	0.06269393595383299
22	0.6790531973275151	0.0763097866723399	0.061556417434235455
23	0.6788857892175738	0.07762627126637756	0.06235338521103506
24	0.6791887021320235	0.07730446332529872	0.06181895571339713
25	0.6820761996410183	0.07680543812742024	0.06150061889964209
26	0.6823589514475619	0.07715485360408769	0.06122013619966726
27	0.6846495634847127	0.07517891319535275	0.06092994164752141
28	0.6843122805515763	0.07699379790223837	0.061537567008248084
29	0.6841018009712196	0.0766190817384792	0.06098588243211467
30	0.6861472055084615	0.07566249186992158	0.060765335228395556
31	0.6860803226658873	0.07627024282809547	0.060751643449870385
32	0.6875554546341669	0.07569401175382473	0.06071914374054399
33	0.6888311399019547	0.07558339524560076	0.060447339662160204
34	0.6865387641982447	0.07672488463552207	0.0608624876156213
35	0.6895996585337929	0.07556728261240908	0.05996222900998893
36	0.6889172974603298	0.07632687468243782	0.06051987207156306
37	0.6891196507449555	0.07594221920783761	0.06033317773991187
38	0.6894891958658742	0.07534942796889699	0.06033080809541813
39	0.6903293498527965	0.07611399608264247	0.06004725756124063
40	0.6905526111520267	0.07604501112628692	0.06007911756905282
41	0.6915470098601013	0.07565991787839688	0.060089788153763535
42	0.6917819029938361	0.07585346368627877	0.05973799116860257
43	0.6911235512965527	0.07583143336125797	0.0598496508052242
44	0.6918457066035457	0.075342478956678	0.06011993784681524
45	0.6913577698787461	0.07615454953926848	0.06031928128608299
46	0.6929845115903884	0.07581835283222171	0.05967903156789445
47	0.6927325898447453	0.07542871813569507	0.05985061138942638
48	0.6934891394442272	0.07550824327455319	0.05967444281721601
49	0.6932798699775137	0.0755615133789519	0.05973091650129814
50	0.6939132101634435	0.07524432943700388	0.059758191990897935
51	0.6936752528866844	0.07529108795163741	0.059700316452340085
52	0.6938415056757126	0.075531961009348	0.05967805011474281
53	0.6943864517693629	0.07535183183954886	0.05953121223116875
54	0.6946068373207794	0.07532379120986032	0.059495268767342124

};
\addlegendentry{Turkers}

\addplot[mark=none, blue, dashed, thick, samples=2, domain=-1:55] {0.6981687724107656};
\addlegendentry{three journalists}

\draw [blue, fill=blue, opacity=0.1] (axis cs:-1,0.6071334714630847) rectangle (axis cs:55,0.7685887889952603);

\addplot[mark=none, green, dashed, thick, samples=2, domain=-1:55] {0.6033601288629524};
\addlegendentry{one journalist}

\draw [green, fill=green, opacity=0.1] (axis cs:-1,0.49990125692341403) rectangle (axis cs:55,0.6887012303041324);

\addplot[mark=*, mark options={scale=.8}, blue, thick]
    table[]{
        54.0	0.6981687724107656
    };

\addplot[mark=none, blue, dashed, thick, samples=2, domain=-1:55] coordinates {(54.0,0.6981687724107656) (54.0,0)};

\addplot[mark=*, mark options={scale=.8}, green, thick]
    table[]{
        5.766977143786355	0.6033601288629524
    };

\addplot[mark=none, green, dashed, thick, samples=2, domain=-1:55] coordinates {(5.766977143786355,0.6033601288629524) (5.766977143786355,0)};


\end{axis}
\end{tikzpicture}
     \begin{tikzpicture}
\sffamily
\begin{axis}[
title = {C2: Independent},
title style={align=center,yshift=-.1in},
legend style={font=\scriptsize,
	nodes={scale=1, transform shape},
	at={(0.0,1)},
	anchor=north west,
	draw=none,
	fill=none},
legend cell align={left},
width = 2.0in, height = 2.5in,
ylabel near ticks,
xlabel near ticks,
every tick label/.append style={font=\scriptsize},
xmin=0,xmax=55,ymin=0.4,ymax=0.9,
xtick={0, 5.185264717019683, 14, 21, 28, 35, 42, 54.0},
tick pos=left,
xlabel={\small Number of raters},
xlabel style = {yshift=0.05in},
yticklabel style={
		/pgf/number format/fixed,
		/pgf/number format/precision=5
},
scaled y ticks=false
]

\addplot[solid, mark=o, mark options={scale=.3}, black]
plot [error bars/.cd, y dir = both, y explicit]
table[y error index=2]{
1	0.42684857982308516	0.07749668294082601	0.06431005097565495
2	0.5068495348485927	0.0802840616346987	0.06786977495298807
3	0.5481013867161111	0.08322033444415189	0.06658301449489501
4	0.5845818708556019	0.08257994036277838	0.06599861545983876
5	0.5993684935068303	0.08306642205691339	0.06652253636191852
6	0.6209140712095618	0.08319663841872338	0.06483655394541044
7	0.6263002164141294	0.08419034525599733	0.06488420937847383
8	0.6362273077729212	0.08271595309786417	0.06364826117425004
9	0.6409590784916529	0.08126643997703176	0.06354650885515079
10	0.6478573352833028	0.08172471825568406	0.06327863604853445
11	0.6514612848150737	0.07991570085635613	0.06369624209339875
12	0.6551498744439138	0.07961649534594928	0.06254153110668548
13	0.66164600814355	0.0797514471533749	0.06214637308808579
14	0.6633954332761866	0.07968853949641008	0.06261723790734974
15	0.66486839099677	0.07884467810863893	0.06178040150960684
16	0.6686894808806788	0.07913486190768715	0.06114069347309259
17	0.6702242062121019	0.07822376426201083	0.061158047892202805
18	0.6727216618470129	0.07875030248826875	0.06094478358854638
19	0.6728821383948681	0.07936348444736674	0.0616923410501804
20	0.6741022396885262	0.07849621324440614	0.06138300165920452
21	0.6746253298770363	0.079103610488099	0.060979644998400695
22	0.6756160978421774	0.07843313986749845	0.06109365640609221
23	0.6788949293782099	0.07849241448770894	0.060161155992837334
24	0.6800007574820492	0.07942389064404265	0.060606629949885193
25	0.6799856228184119	0.07942125418311352	0.06039235691974143
26	0.6807268419222069	0.0780649931007712	0.061051653393781824
27	0.6827940212782928	0.07880977174685466	0.060143681010773054
28	0.6817012933689672	0.07939138827206227	0.060523754893043624
29	0.6833681324948679	0.07799667283997391	0.0608549196405298
30	0.6853784615161266	0.0787209428320349	0.06010801691204748
31	0.6853971418735553	0.07829478425839442	0.05966266700738576
32	0.6856043720063552	0.07876918316325499	0.061031972311317406
33	0.6871128330178706	0.07883668227889862	0.060294901175249604
34	0.6864982091341965	0.07815094634271702	0.06041175988949321
35	0.6873716745810353	0.07787510673797571	0.060344204948130864
36	0.6879598635964693	0.07833972451366833	0.06045026625217731
37	0.6874569393387412	0.0777640843671642	0.06061550055471132
38	0.6887525856628832	0.07832778795481932	0.06023902501128786
39	0.6889143435278541	0.07811635501311731	0.060544986589557204
40	0.6890873377108513	0.07844458473959548	0.06040309981905123
41	0.6898076045867916	0.07823980529757613	0.06033348896595925
42	0.6898341889661264	0.07853502778408028	0.0605966039117366
43	0.6897230665830735	0.07847624153153676	0.060390939892355444
44	0.690712794035319	0.07804698849358749	0.0603332590819875
45	0.690561044660231	0.07819605890208159	0.060498976456556
46	0.6907592452390768	0.07830908404067405	0.06053276871462343
47	0.6918037872077362	0.07777884520466949	0.06036424288022035
48	0.6915631668352895	0.0779023827379961	0.06027536354959351
49	0.6920362482573208	0.0778302222548396	0.06013519002448653
50	0.6918758999973026	0.07782420859640993	0.060381995206482686
51	0.6925744302939186	0.07808397824164637	0.060085587503857396
52	0.6924655027411657	0.07808095122924308	0.060135241347229584
53	0.6927008548574962	0.07801016788106785	0.06009947902503632
54	0.6929100850096386	0.077999957909202	0.06006543582482815

};
\addlegendentry{Turkers}

\addplot[mark=none, blue, dashed, thick, samples=2, domain=-1:55] {0.6981687724107656};
\addlegendentry{three journalists}

\draw [blue, fill=blue, opacity=0.1] (axis cs:-1,0.6071334714630847) rectangle (axis cs:55,0.7685887889952603);

\addplot[mark=none, green, dashed, thick, samples=2, domain=-1:55] {0.6033601288629524};
\addlegendentry{one journalist}

\draw [green, fill=green, opacity=0.1] (axis cs:-1,0.49990125692341403) rectangle (axis cs:55,0.6887012303041324);

\addplot[mark=*, mark options={scale=.8}, blue, thick]
    table[]{
        54.0	0.6981687724107656
    };

\addplot[mark=none, blue, dashed, thick, samples=2, domain=-1:55] coordinates {(54.0,0.6981687724107656) (54.0,0)};

\addplot[mark=*, mark options={scale=.8}, green, thick]
    table[]{
        5.185264717019683	0.6033601288629524
    };

\addplot[mark=none, green, dashed, thick, samples=2, domain=-1:55] coordinates {(5.185264717019683,0.6033601288629524) (5.185264717019683,0)};


\end{axis}
\end{tikzpicture}
    \begin{tikzpicture}
\sffamily
\begin{axis}[
title = {C3: Collective},
title style={align=center,yshift=-.1in},
legend style={font=\scriptsize,
	nodes={scale=1, transform shape},
	at={(0.0,1)},
	anchor=north west,
	draw=none,
	fill=none},
legend cell align={left},
width = 2.0in, height = 2.5in,
ylabel near ticks,
xlabel near ticks,
every tick label/.append style={font=\scriptsize},
xmin=0,xmax=55,ymin=0.4,ymax=0.9,
xtick={0, 7.219838368297723, 14, 21, 28, 35, 42, 54.0},
xlabel={\small Number of raters},
xlabel style = {yshift=0.05in},
yticklabel style={
		/pgf/number format/fixed,
		/pgf/number format/precision=5
},
scaled y ticks=false
]

\addplot[solid, mark=o, mark options={scale=.3}, black]
plot [error bars/.cd, y dir = both, y explicit]
table[y error index=2]{
1	0.4156404808420431	0.09134997833150227	0.07339144050607482
2	0.4828549587361324	0.10345661801834116	0.08061929904641613
3	0.5368584376238456	0.10407515905613424	0.08408476495310191
4	0.5619888776199194	0.10540917955939744	0.08265199334650164
5	0.5820671074189773	0.10428340566973127	0.08056579343308878
6	0.5932592788756524	0.10521483737298937	0.08042881816826641
7	0.6005520671189203	0.10447219528435575	0.07910284600508788
8	0.6133253685484016	0.10215808094847711	0.07742345071744905
9	0.6127905953887173	0.10396194580405638	0.0783826103144395
10	0.6166699051479967	0.10353587703222722	0.07765867273840676
11	0.6258198069839063	0.10519479017552313	0.07601309458914329
12	0.625568279464851	0.10413067604077941	0.07637111592973711
13	0.6278479760868257	0.10508714653851248	0.0771263614575699
14	0.6328417612031698	0.10424550817890665	0.0758164061826544
15	0.6353830127364054	0.10437993568488024	0.07567508427516723
16	0.635205364829162	0.10533423608225057	0.0756331069423345
17	0.6387799477280922	0.10403660831846895	0.07485589071876997
18	0.6393214473721093	0.10403051495241256	0.07490632752319082
19	0.639993023332003	0.10350227079799679	0.07525332116726313
20	0.6430894368983309	0.10325085147067914	0.07452810336078364
21	0.6432285860675373	0.1047816686870432	0.07467246479919376
22	0.6445945222597025	0.10406519979889772	0.07471970586532373
23	0.6458774003094032	0.10356735079595791	0.07383527249367805
24	0.6477090518612022	0.10367602682833876	0.07404117361319118
25	0.6466590144293417	0.10345670558941367	0.07476936114732535
26	0.6488993129083821	0.10358441123313689	0.07420666833245226
27	0.6484394771803514	0.10380844964796543	0.07434972982858168
28	0.6492960028850916	0.10356352096863375	0.0741311572285982
29	0.6506405756234096	0.103728232284088	0.07322517952961993
30	0.6507612603785895	0.10381614565754327	0.07397081397527183
31	0.6509279124727826	0.10376026175375652	0.07367239240214984
32	0.6517089914959787	0.10318106350848655	0.07404288163454542
33	0.6517613594102659	0.10299044294017323	0.0736619673701232
34	0.652478968870127	0.10408053549280638	0.07351192303858833
35	0.6537273970131006	0.10327904051837067	0.0737217347329443
36	0.6552427569563377	0.10283234587644885	0.0731580726570229
37	0.6538872412161166	0.10353537849201577	0.07365317596104626
38	0.6542776560483194	0.10325894910853362	0.07305722238151191
39	0.6544389104445478	0.10286892185114604	0.07323526527115432
40	0.6551876971246517	0.1029665951556673	0.07325914823313195
41	0.6556584948345549	0.10303467425580615	0.0732900271710708
42	0.6559900590442321	0.1029764299175383	0.07310037826417581
43	0.6552304941095345	0.10330045865616666	0.0732550256300607
44	0.6561541974726492	0.10337773612907797	0.07333741392497095
45	0.656407004910404	0.10328340737991415	0.0732127243513695
46	0.6564242757434993	0.10371620186036046	0.07326205261989249
47	0.6572878907986974	0.10318320148099003	0.07311650818414495
48	0.65685601859547	0.10319525221930981	0.07317942858140603
49	0.6569093945229431	0.10332264662700597	0.07326344300214027
50	0.6574646365186656	0.10305563333646106	0.07306919529051614
51	0.6575348761830974	0.10324905127857542	0.0730049820976576
52	0.6578580730549457	0.10316249177886228	0.07305121268732706
53	0.6579337978414294	0.10321442336619813	0.07305261317331879
54	0.6581145028074837	0.10320339626487096	0.07304423722554976

};
\addlegendentry{Turkers}

\addplot[mark=none, blue, dashed, thick, samples=2, domain=-1:55] {0.6981687724107656};
\addlegendentry{three journalists}

\draw [blue, fill=blue, opacity=0.1] (axis cs:-1,0.6071334714630847) rectangle (axis cs:55,0.7685887889952603);

\addplot[mark=none, green, dashed, thick, samples=2, domain=-1:55] {0.6033601288629524};
\addlegendentry{one journalist}

\draw [green, fill=green, opacity=0.1] (axis cs:-1,0.49990125692341403) rectangle (axis cs:55,0.6887012303041324);

\addplot[mark=*, mark options={scale=.8}, blue, thick]
    table[]{
        54.0	0.6981687724107656
    };

\addplot[mark=none, blue, dashed, thick, samples=2, domain=-1:55] coordinates {(54.0,0.6981687724107656) (54.0,0)};

\addplot[mark=*, mark options={scale=.8}, green, thick]
    table[]{
        7.219838368297723	0.6033601288629524
    };

\addplot[mark=none, green, dashed, thick, samples=2, domain=-1:55] coordinates {(7.219838368297723,0.6033601288629524) (7.219838368297723,0)};


\end{axis}
\end{tikzpicture}
    \caption{Power curves for the three conditions for only the 109 political items. The x-axis is the number of turkers. The y-axis is the correlation of the mean of $k$ turkers' ratings with a journalist's rating. The green horizontal lines show the correlation of a randomly selected journalist's rating for each item with a held-out journalist (0.60). The blue lines shows the correlation of the mean of three journalists with a held-out journalist (0.69).}
    \label{fig:survey-equivalence-political-items}
\end{figure*}
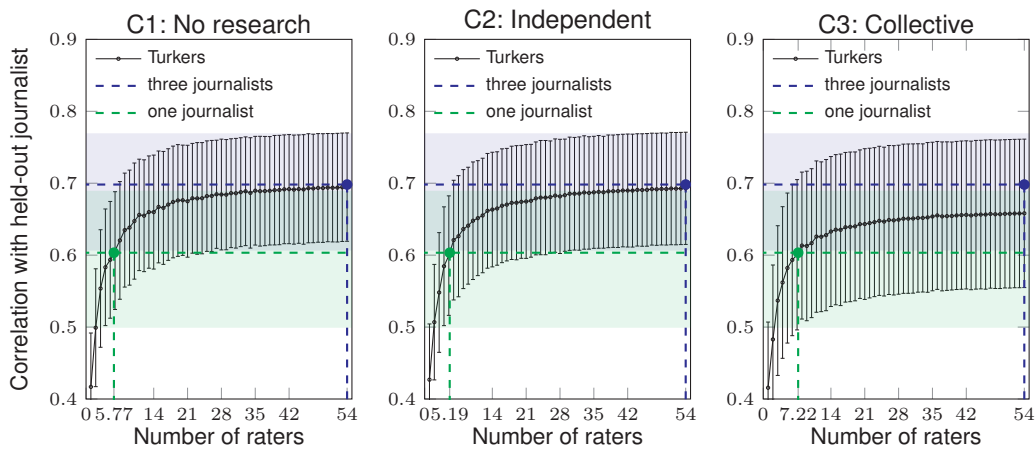

Since the results are somewhat different for the political items, Figure~\ref{fig:survey-equivalence-political-items} plots the power curves and survey equivalence values for just those items.

\begin{table}
\centering
\footnotesize
\caption{Percent of Raters Who Reported Not Enough Information to Make a Judgment, per Item}
\label{tab:not-enough-info}
\begin{tabular}{@{}lrrrc@{}}
 & \begin{tabular}[r]{@{}r@{}r@{}}\textbf{C1}\\\textbf{No}\\\textbf{Research}\end{tabular} 
 & \begin{tabular}[r]{@{}r@{}r@{}}\textbf{C2} \\ \textbf{Individual}\\\textbf{Research}\end{tabular} 
 & \begin{tabular}[r]{@{}r@{}r@{}}\textbf{C3} \\ \textbf{Collective}\\\textbf{Research}\end{tabular} 
 &  \\ [1ex]
 \midrule
Mean & 1.5\%  & 3.0\%  &  1.4\% \\ [1ex]
Median & 0  & 1.9\%  & 0  \\ [1ex] 
Max  & 7.4\% & 22.2\% & 35.7\% \\ [1ex] 
\midrule
\end{tabular}
\label{tab:ratings_per_turker}
\end{table}

\begin{table}
\centering
\small
\caption{Reliability of Comparisons Between Conditions. Cells Show Proportion of Bootstrap Item Samples.}
\vspace{5px}
\label{tab:reliability-of-comparisons}
\begin{tabular}{@{}lrrrc@{}}
number of lay raters & \textbf{C2 $>$ C1} & \textbf{C3 $>$ C1} & \textbf{C3 $>$ C2} &  \\ [1ex]
 \midrule
1 & 1.00 & 1.00 & 0.71 \\ [1ex]
2 & 1.00 & 1.00 & 0.28 \\ [1ex]
3 & 1.00 & 1.00 & 0.24 \\ [1ex]
4 & 1.00 & 1.00 & 0.05 \\ [1ex]
5 & 1.00 & 1.00 & 0.02 \\ [1ex]
6 & 1.00 & 1.00 & 0.00 \\ [1ex]
7 & 1.00 & 1.00 & 0.00 \\ [1ex]
8 & 1.00 & 0.99 & 0.01 \\ [1ex]
9 & 1.00 & 0.98 & 0.00 \\ [1ex]
10 & 1.00 & 0.97 & 0.00 \\ [1ex]
12 & 1.00 & 0.98 & 0.00 \\ [1ex]
14 & 1.00 & 0.98 & 0.00 \\ [1ex]
16 & 1.00 & 0.96 & 0.00 \\ [1ex]
18 & 1.00 & 0.96 & 0.00 \\ [1ex]
20 & 1.00 & 0.96 & 0.00 \\ [1ex]
22 & 1.00 & 0.95 & 0.00 \\ [1ex]
24 & 1.00 & 0.96 & 0.00 \\ [1ex]
26 & 1.00 & 0.95 & 0.00 \\ [1ex]
28 & 1.00 & 0.95 & 0.00 \\ [1ex]
30 & 1.00 & 0.95 & 0.00 \\ [1ex]
32 & 1.00 & 0.94 & 0.00 \\ [1ex]
34 & 1.00 & 0.95 & 0.00 \\ [1ex]
36 & 1.00 & 0.94 & 0.00 \\ [1ex]
38 & 1.00 & 0.94 & 0.00 \\ [1ex]
40 & 1.00 & 0.94 & 0.00 \\ [1ex]
42 & 1.00 & 0.94 & 0.00 \\ [1ex]
44 & 1.00 & 0.94 & 0.00 \\ [1ex]
46 & 1.00 & 0.94 & 0.00 \\ [1ex]
48 & 1.00 & 0.94 & 0.00 \\ [1ex]
50 & 1.00 & 0.94 & 0.00 \\ [1ex]
52 & 1.00 & 0.94 & 0.00 \\ [1ex]
54 & 1.00 & 0.94 & 0.00 \\ [1ex]
\midrule
\end{tabular}
\end{table}

\begin{table*}[!]
\centering
\footnotesize
\caption{Survey equivalence confidence intervals. How many lay raters provide the same power to predict a journalist as a panel of m journalists? Confidence intervals based on covering 95\% of the survey equivalence values from 500 bootstrap samples of articles.}
\vspace{5px}
\label{tab:seq-cis}
\begin{tabular}{@{}lrrrc@{}}
m journalists & \textbf{C1:No Research} & \textbf{C2: Individual Research} & \textbf{C3: Collective Research} &  \\ [1ex]
 \midrule
1  & (3.92, 24.58) & (2.53, 4.62) & (2.62, 5.94) \\ [1ex]
2  & (12.26, $>$54) & (4.75, 12.14) & (6.01, $>$54) \\ [1ex]
3  & ($>$54, $>$54) & (8.30, 53.42) & (17.38, $>$54) \\ [1ex]
\midrule
\end{tabular}
\end{table*}

\begin{table*}
\centering
\small
\caption{Proportion of bootstrap samples of articles where k lay raters are better than m journalists (i.e., higher correlation with held-out journalist}
\vspace{5px}
\label{tab:reliability-of-equivalences}
\begin{tabular}{@{}l|lll|lll|lll@{}}

k lay & \multicolumn{3}{c}{\textbf{C1: No Research}} & \multicolumn{3}{|c}{\textbf{C2: Individual Research}} & \multicolumn{3}{|c}{\textbf{C3: Collective Research}} \\
raters & \multicolumn{3}{c}{m journalists} & \multicolumn{3}{|c}{m journalists} & \multicolumn{3}{|c}{m journalists}  \\ [1ex]
 & 1 & 2 & 3  & 1 & 2 & 3  & 1 & 2 & 3 	\\ [1ex]
 \midrule
 1  & 0.00 & 0.00 & 0.00 & 0.00 & 0.00 & 0.00 & 0.00 & 0.00 & 0.00 \\ [1ex]
2  & 0.00 & 0.00 & 0.00 & 0.00 & 0.00 & 0.00 & 0.00 & 0.00 & 0.00 \\ [1ex]
3  & 0.00 & 0.00 & 0.00 & 0.24 & 0.00 & 0.00 & 0.16 & 0.00 & 0.00 \\ [1ex]
4  & 0.03 & 0.00 & 0.00 & 0.88 & 0.00 & 0.00 & 0.63 & 0.00 & 0.00 \\ [1ex]
5  & 0.09 & 0.00 & 0.00 & 1.00 & 0.06 & 0.00 & 0.89 & 0.01 & 0.00 \\ [1ex]
6  & 0.19 & 0.00 & 0.00 & 1.00 & 0.29 & 0.00 & 0.98 & 0.03 & 0.00 \\ [1ex]
7  & 0.38 & 0.00 & 0.00 & 1.00 & 0.50 & 0.00 & 0.99 & 0.08 & 0.00 \\ [1ex]
8  & 0.58 & 0.00 & 0.00 & 1.00 & 0.68 & 0.01 & 1.00 & 0.18 & 0.00 \\ [1ex]
9  & 0.67 & 0.00 & 0.00 & 1.00 & 0.83 & 0.05 & 1.00 & 0.21 & 0.00 \\ [1ex]
10  & 0.76 & 0.01 & 0.00 & 1.00 & 0.91 & 0.13 & 1.00 & 0.28 & 0.00 \\ [1ex]
12  & 0.82 & 0.02 & 0.00 & 1.00 & 0.97 & 0.26 & 1.00 & 0.41 & 0.01 \\ [1ex]
14  & 0.89 & 0.04 & 0.00 & 1.00 & 0.99 & 0.44 & 1.00 & 0.57 & 0.01 \\ [1ex]
16  & 0.93 & 0.06 & 0.00 & 1.00 & 1.00 & 0.57 & 1.00 & 0.61 & 0.02 \\ [1ex]
18  & 0.95 & 0.11 & 0.00 & 1.00 & 1.00 & 0.66 & 1.00 & 0.68 & 0.03 \\ [1ex]
20  & 0.96 & 0.11 & 0.00 & 1.00 & 1.00 & 0.72 & 1.00 & 0.73 & 0.04 \\ [1ex]
22  & 0.97 & 0.14 & 0.00 & 1.00 & 1.00 & 0.77 & 1.00 & 0.76 & 0.06 \\ [1ex]
24  & 0.97 & 0.15 & 0.00 & 1.00 & 1.00 & 0.80 & 1.00 & 0.80 & 0.08 \\ [1ex]
26  & 0.98 & 0.19 & 0.00 & 1.00 & 1.00 & 0.83 & 1.00 & 0.83 & 0.09 \\ [1ex]
28  & 0.98 & 0.22 & 0.00 & 1.00 & 1.00 & 0.85 & 1.00 & 0.83 & 0.10 \\ [1ex]
30  & 0.98 & 0.25 & 0.00 & 1.00 & 1.00 & 0.89 & 1.00 & 0.86 & 0.11 \\ [1ex]
32  & 0.98 & 0.27 & 0.01 & 1.00 & 1.00 & 0.91 & 1.00 & 0.87 & 0.12 \\ [1ex]
34  & 0.98 & 0.27 & 0.01 & 1.00 & 1.00 & 0.92 & 1.00 & 0.89 & 0.12 \\ [1ex]
36  & 0.99 & 0.30 & 0.01 & 1.00 & 1.00 & 0.94 & 1.00 & 0.90 & 0.13 \\ [1ex]
38  & 0.99 & 0.32 & 0.01 & 1.00 & 1.00 & 0.95 & 1.00 & 0.90 & 0.14 \\ [1ex]
40  & 0.99 & 0.32 & 0.01 & 1.00 & 1.00 & 0.96 & 1.00 & 0.90 & 0.14 \\ [1ex]
42  & 0.99 & 0.33 & 0.01 & 1.00 & 1.00 & 0.96 & 1.00 & 0.91 & 0.15 \\ [1ex]
44  & 0.99 & 0.34 & 0.01 & 1.00 & 1.00 & 0.96 & 1.00 & 0.91 & 0.15 \\ [1ex]
46  & 0.99 & 0.35 & 0.01 & 1.00 & 1.00 & 0.96 & 1.00 & 0.91 & 0.16 \\ [1ex]
48  & 0.99 & 0.36 & 0.01 & 1.00 & 1.00 & 0.96 & 1.00 & 0.92 & 0.17 \\ [1ex]
50  & 0.99 & 0.37 & 0.01 & 1.00 & 1.00 & 0.97 & 1.00 & 0.92 & 0.17 \\ [1ex]
52  & 0.99 & 0.37 & 0.01 & 1.00 & 1.00 & 0.97 & 1.00 & 0.93 & 0.18 \\ [1ex]
54  & 0.99 & 0.39 & 0.01 & 1.00 & 1.00 & 0.97 & 1.00 & 0.93 & 0.19 \\ [1ex]
\midrule
\end{tabular}
\end{table*}

\begin{table*}
\centering
\small
\begin{tabular}{lrrr}
 & \textbf{C1: No Research} & \textbf{C2: Individual Research} & \textbf{C3: Collective Research}  \\ [1ex]
 \midrule
Liberal-Conservative Correlation & .80 & .88 & .91  \\ [1ex] 
Correlation of one lay rater with one journalist  & 0.47  & 0.53 & 0.54 \\ [1ex] 
Lay raters equivalent to one journalist & 7.85  & 3.80 & 4.29 \\ [1ex] 
Lay raters equivalent to two journalists & $>$54  & 9.06 & 25.92 \\ [1ex] 
Lay raters equivalent to three journalists & $>$54  & 25.11 & $>$54 \\ [1ex] 
Liberal raters equivalent to one journalist & 3.86  & 2.48 & 2.33 \\ [1ex] 
Liberal raters equivalent to two journalists & 13.29  & 5.00 & 6.26 \\ [1ex] 
Liberal raters equivalent to three journalists & $>$18  & 10.01 & $>$18 \\ [1ex] 
Conservative raters equivalent to one journalist & $>$18  & 6.78 & $>$18 \\ [1ex] 
Conservative raters equivalent to two journalists & $>$18  & $>$18 & $>$18 \\ [1ex] 
Conservative raters equivalent to three journalists & $>$18  & $>$18 & $>$18 \\ [1ex] 
\end{tabular}
\vspace{4pt}
\caption{Summary of results for article collection one only, using our four journalist ratings.}
\label{tab:results-summary-mit}
\end{table*}

\begin{table*}
\centering
\small
\begin{tabular}{lrrr}
 & \textbf{C1: No Research} & \textbf{C2: Individual Research} & \textbf{C3: Collective Research}  \\ [1ex]
 \midrule
Liberal-Conservative Correlation & .80 & .88 & .91  \\ [1ex] 
Correlation of one lay rater with one journalist  & 0.45  & 0.50 & 0.50 \\ [1ex] 
Lay raters equivalent to one journalist & 5.46  & 3.55 & 4.38 \\ [1ex] 
Lay raters equivalent to two journalists & $>$54  & 16.22 & $>$54 \\ [1ex] 
Liberal raters equivalent to one journalist & 4.70  & 2.67 & 2.78 \\ [1ex] 
Liberal raters equivalent to two journalists & $>$18  & 13.60 & $>$18 \\ [1ex] 
Conservative raters equivalent to one journalist & 7.95  & 4.78 & 6.66 \\ [1ex] 
Conservative raters equivalent to two journalists & $>$18  & $>$18 & $>$18 \\ [1ex] 
\end{tabular}
\vspace{4pt}
\caption{Summary of results for article collection one only, using the journalist ratings from
~\protect\cite{allen2020scaling}. 
Note that it is only only possible to compute equivalences to panels of one or two journalists, because only three journalist ratings were available and one journalist always has to be held out as the reference rater.}
\label{tab:results-summary-mit-mit-journalists}
\end{table*}

\begin{table*}
\centering
\begin{tabular}{lrrr}
 & \textbf{C1} & \textbf{C2} & \textbf{C3} \\ [1ex]
 \midrule
Liberal-Conservative Correlation & .83 & .83 & .85  \\ [1ex] 
Correlation of one lay rater with one journalist  & 0.46  & 0.52 & 0.53 \\ [1ex] 
Lay raters equivalent to one journalist & 3.83  & 2.87 & 3.81 \\ [1ex] 
Lay raters equivalent to two journalists & 14.22  & 5.42 & 9.66 \\ [1ex] 
Lay raters equivalent to three journalists & $>$54  & 10.30 & 40.02 \\ [1ex] 
Liberal raters equivalent to one journalist & 2.26  & 1.77 & 2.38 \\ [1ex] 
Liberal raters equivalent to two journalists & 4.77  & 3.06 & 4.86 \\ [1ex] 
Liberal raters equivalent to three journalists & 10.57 & 5.03 & 9.29 \\ [1ex] 
Conservative raters equivalent to one journalist & $>$18  & 4.65 & 7.48 \\ [1ex] 
Conservative raters equivalent to two journalists & $>$18  & 13.99 & $>$18 \\ [1ex] 
Conservative raters equivalent to three journalists & $>$18  & $>$18 & $>$18 \\ [1ex] 
\end{tabular}
\vspace{4pt}
\caption{Summary of results for article collection two only, using our four journalist ratings.}
\label{tab:results-summary-nyu}
\end{table*}

\begin{table*}
\centering
\small
\begin{tabular}{lrrr}
 & \textbf{C1: No Research} & \textbf{C2: Individual Research} & \textbf{C3: Collective Research}  \\ [1ex]
 \midrule
Liberal-Conservative Correlation & .69 & .73 & .79  \\ [1ex] 
Correlation of one lay rater with one journalist  & 0.28  & 0.33 & 0.33 \\ [1ex] 
Lay raters equivalent to one journalist & 16.49  & 3.86 & 4.60 \\ [1ex] 
Lay raters equivalent to two journalists & $>$54  & 7.71 & 13.86 \\ [1ex] 
Lay raters equivalent to three journalists & $>$54  & 18.82 & $>$54 \\ [1ex] 
Liberal raters equivalent to one journalist & 4.00  & 2.12 & 2.41 \\ [1ex] 
Liberal raters equivalent to two journalists & 10.83  & 3.69 & 4.69 \\ [1ex] 
Conservative raters equivalent to one journalist & $>$18  & 6.57 & 10.21 \\ [1ex] 
Conservative raters equivalent to two journalists & $>$18  & $>$18 & $>$18 \\ [1ex] 
\end{tabular}
\vspace{4pt}
\caption{Summary of results for only the 98 articles from collection two that came from non-mainstream sites.}
\label{tab:results-summary-nyu-fringe-items}
\end{table*}

\begin{figure*}
    \centering
   \pgfplotsset{
    small,
    legend style={
        at={(0.01,0.01)},
        anchor=south west,
    },
   }%
   \pgfdeclareplotmark{fat-}
{%
  \pgfsetlinewidth{0.1pt}
   \pgfsetplotmarksize{.15ex}
  \pgfpathmoveto{\pgfqpoint{\pgfplotmarksize}{0pt}}%
  \pgfpathlineto{\pgfqpoint{-\pgfplotmarksize}{0pt}}%
  \pgfusepathqstroke
}%
\pgfplotsset{
/pgfplots/error bars/error bar style={ultra thin},
/pgfplots/error bars/error mark={fat-}
}
    \begin{tikzpicture}
\sffamily
\begin{axis}[
title = {C1: No research},
title style={align=center,yshift=-.1in},
legend style={font=\scriptsize,
	nodes={scale=1, transform shape},
	at={(0.0,1)},
	anchor=north west,
	draw=none,
	fill=none},
legend cell align={left},
width = 2.0in, height = 2.5in,
ylabel near ticks,
ylabel = {\small Correlation with held-out journalist},
xlabel near ticks,
every tick label/.append style={font=\scriptsize},
xmin=0,xmax=55,ymin=0.4,ymax=0.9,
xtick={16.490929354566056, 54},
tick pos=left,
xlabel={\small Number of raters},
xlabel style = {yshift=0.05in},
yticklabel style={
		/pgf/number format/fixed,
		/pgf/number format/precision=5
},
scaled y ticks=false
]

\addplot[solid, mark=o, mark options={scale=.3}, black]
plot [error bars/.cd, y dir = both, y explicit]
table[y error index=2]{
1	0.2768567623207157	0.07903103046110985	0.08524516882649297
2	0.3437902516183604	0.0944083091262973	0.09735190493532725
3	0.37567011111567183	0.09987820614446041	0.10048218826807653
4	0.404155830868697	0.10730965325438646	0.10217253319789088
5	0.4191098231264202	0.11198940725013334	0.10578011321968989
6	0.4332652091774672	0.11163366697001076	0.1003307958640457
7	0.43722963033280604	0.11569934032747098	0.1013680536479385
8	0.443981416175344	0.11462732904968714	0.10215161251479982
9	0.45171507025509267	0.11639122194042356	0.10252597559974569
10	0.448769206691209	0.11896445865817595	0.10164659532265263
11	0.45882045920669223	0.11800494180441179	0.10231058608733995
12	0.461066201288731	0.11695074964145835	0.10120291827405353
13	0.4658658395491316	0.11980976997281562	0.10235684295870712
14	0.46690606422577247	0.12202087038394793	0.10097270069281317
15	0.47287183567333463	0.12020673732119685	0.10074357229848102
16	0.4695417715554763	0.12183899656279346	0.10204620979425638
17	0.47701083376582687	0.12187986755299418	0.10021113018424244
18	0.4743640913170244	0.12188429441111365	0.10149994136479484
19	0.47545302824196334	0.12390544854380381	0.10147671333588776
20	0.47976905422389543	0.12256536619634756	0.10126974145004974
21	0.4809897008954017	0.12333921232791073	0.10111079750979868
22	0.4785219240847709	0.12238366344974916	0.10187386372470286
23	0.48446987020541016	0.12152070899901313	0.10044756572503899
24	0.48615077850305843	0.12193421562299772	0.1011536434648394
25	0.48363073359718906	0.1231905055721571	0.10155524239607444
26	0.4843690673639129	0.12304996446327959	0.10154992483372777
27	0.48526958598629194	0.12411634645749009	0.10170775166134277
28	0.48813245689647794	0.12267423933119559	0.10139945804851125
29	0.486372962295717	0.12358027815971528	0.10188770879898579
30	0.48887578352935046	0.12185407111492103	0.10053629237231965
31	0.48648413465097057	0.12385560936835327	0.10135477610852095
32	0.48976993092170906	0.12383384889075372	0.10114543357866945
33	0.4891536515161974	0.12382148452101482	0.10091818023756549
34	0.49098160075960734	0.12286866780112582	0.1018330401006588
35	0.4911995236096396	0.12408216583245807	0.10136413490990254
36	0.49121979301059054	0.12325109583600236	0.10103010409523205
37	0.4905002071523119	0.12377343424597137	0.10185840640226518
38	0.48998951598508483	0.1245060384427652	0.10159710780086106
39	0.4924982117969334	0.1238182449587592	0.10074872154743103
40	0.49234797160014915	0.12353409163621276	0.10123970529834614
41	0.49183668276264486	0.12366155451342825	0.10098619934791964
42	0.49335771116483385	0.12470831851661968	0.10154010241149891
43	0.4929215049946508	0.12434433748272661	0.1007860719661885
44	0.4926536866106829	0.12389671865800073	0.10126778028620276
45	0.49462602415137646	0.12389342077893539	0.10067516431107681
46	0.4933461164883069	0.12465734244560861	0.10148823359005871
47	0.49438267265325436	0.1245588119340359	0.10080057831951234
48	0.494330326572601	0.12466930052187752	0.1013627544300485
49	0.4957849330635231	0.12379439732554903	0.10093624924728378
50	0.4955344981046666	0.12368589337449237	0.10129521465099639
51	0.49548144112351855	0.12403708068554331	0.10083465431106015
52	0.4957758790446611	0.1239115293936286	0.10089982577833734
53	0.4960348493059341	0.12405783332270232	0.10104532559645563
54	0.49623411254068805	0.12408085867141039	0.1010386733238412

};
\addlegendentry{Turkers}

\addplot[mark=none, blue, dashed, thick, samples=2, domain=-1:55] {0.5723571619850512};
\addlegendentry{three journalists}

\draw [blue, fill=blue, opacity=0.1] (axis cs:-1,0.4498635995199335) rectangle (axis cs:55,0.6754711088720982);

\addplot[mark=none, green, dashed, thick, samples=2, domain=-1:55] {0.4732085534456174};
\addlegendentry{one journalist}

\draw [green, fill=green, opacity=0.1] (axis cs:-1,0.35226481091883033) rectangle (axis cs:55,0.5914083104837113);

\addplot[mark=*, mark options={scale=.8}, blue, thick]
    table[]{
        54.0	0.5723571619850512
    };

\addplot[mark=none, blue, dashed, thick, samples=2, domain=-1:55] coordinates {(54.0,0.5723571619850512) (54.0,0)};

\addplot[mark=*, mark options={scale=.8}, green, thick]
    table[]{
        16.490929354566056	0.4732085534456174
    };

\addplot[mark=none, green, dashed, thick, samples=2, domain=-1:55] coordinates {(16.490929354566056,0.4732085534456174) (16.490929354566056,0)};

\end{axis}
\end{tikzpicture}
     \begin{tikzpicture}
\sffamily
\begin{axis}[
title = {C2: Independent},
title style={align=center,yshift=-.1in},
legend style={font=\scriptsize,
	nodes={scale=1, transform shape},
	at={(0.0,1)},
	anchor=north west,
	draw=none,
	fill=none},
legend cell align={left},
width = 2.0in, height = 2.5in,
ylabel near ticks,
xlabel near ticks,
every tick label/.append style={font=\scriptsize},
xmin=0,xmax=55,ymin=0.4,ymax=0.9,
xtick={3.860137747515382, 18.822232967367984, 54},
tick pos=left,
xlabel={\small Number of raters},
xlabel style = {yshift=0.05in},
yticklabel style={
		/pgf/number format/fixed,
		/pgf/number format/precision=5
},
scaled y ticks=false
]

\addplot[solid, mark=o, mark options={scale=.3}, black]
plot [error bars/.cd, y dir = both, y explicit]
table[y error index=2]{
1	0.3327975320441964	0.08376710583515162	0.08828527970437583
2	0.4002859119455848	0.09242409504661375	0.09784002004013792
3	0.44336430330195076	0.09595011016839344	0.1020813271452346
4	0.4780613622670707	0.09647751543068045	0.10076860007821309
5	0.4976740872326162	0.09763227859967899	0.10191152361877681
6	0.5116327441250524	0.09794737745807525	0.10389056379015438
7	0.5201165816401937	0.09769587340145525	0.1054687331089793
8	0.5335523352617246	0.0969542071775975	0.10407162619071353
9	0.5418965630563449	0.09596665964628381	0.103262066106521
10	0.5444009017869981	0.09554956289210803	0.10423757668591205
11	0.5528301495237603	0.0950430500914744	0.10335743248340779
12	0.5543514756143125	0.09552548300740454	0.10472692815608109
13	0.5611843059764324	0.09422164322998561	0.10283892563092933
14	0.5597250537391799	0.09571443619148057	0.10466564997407224
15	0.5683962631964455	0.09409888408940775	0.10260218683165889
16	0.5708088873291548	0.09502275269518623	0.10094050367647456
17	0.5662881128834957	0.09531117098592312	0.10338604594069312
18	0.5690432688993573	0.09428800096070084	0.10193033752839042
19	0.5730736267385739	0.09458635350665356	0.10159280677697413
20	0.5771923504924495	0.09433020393278174	0.10133015963087477
21	0.5781438948190395	0.09484438873963719	0.10093619606349435
22	0.5777012146088043	0.09434356413591366	0.10143112888402495
23	0.5794947745798388	0.09521209574848166	0.10158595209790544
24	0.5827346779355246	0.09416987618768646	0.10001221847457942
25	0.5816685515144829	0.09436392558691098	0.10102663844990412
26	0.584602682037088	0.09384279585224203	0.09990867046237906
27	0.5851579490882661	0.09359234287399737	0.1005567631038371
28	0.585891697472915	0.09356912350908236	0.09967780027320783
29	0.5845549721065784	0.09384331163766846	0.10128870028174652
30	0.5873952937011093	0.0933193436396953	0.10014387423630855
31	0.5882568340921825	0.09259675771151354	0.09976068259727511
32	0.588217008675902	0.0937950425997226	0.10097333304792944
33	0.5903437491480069	0.09313328880431326	0.09940242952988365
34	0.5895793928445363	0.09382279562661722	0.10067895737376842
35	0.5893647514270622	0.09389622238767081	0.10071187613697674
36	0.5904132259381827	0.09375356033519577	0.10033189523328356
37	0.5916798627801255	0.09288529319323702	0.09996590535971561
38	0.5913572387123888	0.09350357306375501	0.10057208767213444
39	0.5921976188408195	0.09290212384759955	0.10058230217049413
40	0.5917211115482298	0.09313852354120367	0.10027880706008818
41	0.5934158496213127	0.09331407987467294	0.09947728598318617
42	0.5939362841502348	0.0927969421505781	0.09956521745383418
43	0.5937701722053792	0.09307630267997535	0.09941213754380163
44	0.5938308196927111	0.09316593124111305	0.09943874035103661
45	0.5946065806261609	0.0928213230016508	0.09909388630658833
46	0.5957056807799823	0.0925377128506335	0.09871920204707552
47	0.5960445406651611	0.09270766360311145	0.09845208752861567
48	0.5960774291530361	0.09249736755202265	0.09856873873691141
49	0.595684062700476	0.09292893481130549	0.09882321741268851
50	0.5956479156115752	0.09283091549496014	0.09893710164619662
51	0.596381117577363	0.09280889850661178	0.09861023766888855
52	0.5966883319067174	0.09286056058709646	0.09872455241378253
53	0.5970839727932467	0.09264570455360677	0.09843664742772584
54	0.5973329614101074	0.0926187993137676	0.09836199533243517

};
\addlegendentry{Turkers}

\addplot[mark=none, blue, dashed, thick, samples=2, domain=-1:55] {0.5723571619850512};
\addlegendentry{three journalists}

\draw [blue, fill=blue, opacity=0.1] (axis cs:-1,0.4498635995199335) rectangle (axis cs:55,0.6754711088720982);

\addplot[mark=none, green, dashed, thick, samples=2, domain=-1:55] {0.4732085534456174};
\addlegendentry{one journalist}

\draw [green, fill=green, opacity=0.1] (axis cs:-1,0.35226481091883033) rectangle (axis cs:55,0.5914083104837113);

\addplot[mark=*, mark options={scale=.8}, blue, thick]
    table[]{
        18.822232967367984	0.5723571619850512
    };

\addplot[mark=none, blue, dashed, thick, samples=2, domain=-1:31] coordinates {(18.822232967367984,0.5723571619850512) (18.822232967367984,0)};

\addplot[mark=*, mark options={scale=.8}, green, thick]
    table[]{
        3.860137747515382	0.4732085534456174
    };

\addplot[mark=none, green, dashed, thick, samples=2, domain=-1:55] coordinates {(3.860137747515382,0.4732085534456174) (3.860137747515382,0)};


\end{axis}
\end{tikzpicture}
    \begin{tikzpicture}
\sffamily
\begin{axis}[
title = {C3: Collective},
title style={align=center,yshift=-.1in},
legend style={font=\scriptsize,
	nodes={scale=1, transform shape},
	at={(0.0,1)},
	anchor=north west,
	draw=none,
	fill=none},
legend cell align={left},
width = 2.0in, height = 2.5in,
ylabel near ticks,
xlabel near ticks,
every tick label/.append style={font=\scriptsize},
xmin=0,xmax=55,ymin=0.4,ymax=0.9,
xtick={4.600919020012779, 54},
tick pos=left,
xlabel={\small Number of raters},
xlabel style = {yshift=0.05in},
yticklabel style={
		/pgf/number format/fixed,
		/pgf/number format/precision=5
},
scaled y ticks=false
]

\addplot[solid, mark=o, mark options={scale=.3}, black]
plot [error bars/.cd, y dir = both, y explicit, ]
table[y error index=2]{
1	0.32991871378318693	0.09717467453661799	0.10035641607415069
2	0.39106700087299634	0.10674453396802364	0.10913218337654829
3	0.4355296745269772	0.11284819765514409	0.10987913246437386
4	0.46037008008448127	0.1115355136970258	0.10918676652664644
5	0.4817348113270755	0.11407397248535994	0.10845338571663482
6	0.4925600752362034	0.11199949637620749	0.10808409403656699
7	0.4973919710305039	0.114508310064104	0.10883959684455319
8	0.5115659768621595	0.10917150329253655	0.10441669695446743
9	0.5137327511104047	0.1115923202723485	0.1051685408549401
10	0.5210689239175338	0.10682186348191125	0.10588067624834019
11	0.5213758932739068	0.11154298756174169	0.10530702972256534
12	0.5260108978059502	0.108872388177774	0.10374222257888632
13	0.528961023171001	0.1096092172429658	0.10714877666110167
14	0.5297250643269255	0.10843521371899656	0.10380716691041492
15	0.5353981188695495	0.10702324329768875	0.10464404655569837
16	0.5393170947113892	0.10767573732964181	0.10358927937846096
17	0.5396434422765567	0.10803179927262535	0.10424609772274951
18	0.5387318843769214	0.10782816701631126	0.10458533043436569
19	0.5414858132770196	0.10652322132159203	0.10421074449110557
20	0.5451187701096514	0.10631603623475272	0.10371020008425813
21	0.5445369256843845	0.10700420116113585	0.10321113741055754
22	0.5442196251957838	0.10781542944020239	0.10369246092126827
23	0.5455738456904867	0.1065496765753775	0.10348815034637149
24	0.5488433712242307	0.10618433444118486	0.10198797357861722
25	0.5501088480884393	0.10617439394453859	0.10165235931627115
26	0.5480183760458243	0.10747259926918784	0.10255261993415687
27	0.5518015268146538	0.10575283426598037	0.10294480009473161
28	0.5519739081385252	0.10576947777246809	0.10191901194518527
29	0.551098893892553	0.1064310504046928	0.10189746545768674
30	0.5520989143563612	0.10672262955693518	0.10188006534169347
31	0.5532088264175277	0.10584096056138947	0.10227882677703626
32	0.5545167362081462	0.10636180929942701	0.1020136917938399
33	0.5539453593285523	0.10593382469798607	0.10213934643416134
34	0.5559291613785847	0.10594962657899842	0.10135644999382787
35	0.5561482918156426	0.10593999920531594	0.10224306827483431
36	0.5559334665902875	0.10635622538170819	0.10167360971861739
37	0.5556666034582356	0.10547080100370498	0.10189997474043344
38	0.5565342948753527	0.10536276032452019	0.10161825186797047
39	0.5566824874693167	0.1057450642710529	0.1020163003178165
40	0.5576032685599375	0.10535660241156064	0.10144393004011876
41	0.5577350986842051	0.10521061492699724	0.10167022275452309
42	0.5574833310758318	0.10548498121649985	0.10195817757340198
43	0.5585332703629754	0.10487748128657493	0.10160416044337806
44	0.5570633597219202	0.1054121522044843	0.10182613450521327
45	0.5582138111382957	0.10565390223999643	0.10139538122846459
46	0.5595474917864743	0.10528914230753472	0.10173388767976099
47	0.5599362764186376	0.10505663560191397	0.10169426431927342
48	0.5598264220405911	0.10549033321696638	0.10166796659729616
49	0.5599736835175986	0.10527294126618647	0.10176155531382336
50	0.5600429958408795	0.10502138464892796	0.1017211717862
51	0.560327696641492	0.10495333560292508	0.10176829958494582
52	0.5606876962416919	0.10492706284638958	0.10161958110918357
53	0.5606220310240579	0.10503544287634292	0.10169172303229768
54	0.5608174040926154	0.10500914924613991	0.10169817578889606

};
\addlegendentry{Turkers}

\addplot[mark=none, blue, dashed, thick, samples=2, domain=-1:55] {0.5723571619850512};
\addlegendentry{three journalists}

\draw [blue, fill=blue, opacity=0.1] (axis cs:-1,0.4498635995199335) rectangle (axis cs:55,0.6754711088720982);

\addplot[mark=none, green, dashed, thick, samples=2, domain=-1:55] {0.4732085534456174};
\addlegendentry{one journalist}

\draw [green, fill=green, opacity=0.1] (axis cs:-1,0.35226481091883033) rectangle (axis cs:55,0.5914083104837113);

\addplot[mark=*, mark options={scale=.8}, blue, thick]
    table[]{
        54.0	0.5723571619850512
    };

\addplot[mark=none, blue, dashed, thick, samples=2, domain=-1:55] coordinates {(54.0,0.5723571619850512) (54.0,0)};

\addplot[mark=*, mark options={scale=.8}, green, thick]
    table[]{
        4.600919020012779	0.4732085534456174
    };

\addplot[mark=none, green, dashed, thick, samples=2, domain=-1:55] coordinates {(4.600919020012779,0.4732085534456174) (4.600919020012779,0)};


\end{axis}
\end{tikzpicture}
    \caption{Power curves for the three conditions for only the 98 articles from collection two that came from non-mainstream sites. The x-axis is the number of turkers. The y-axis is the correlation of the mean of $k$ turkers' ratings with a journalist's rating. The green horizontal lines show the correlation of a randomly selected journalist's rating for each item with a held-out journalist (0.47). The blue lines shows the correlation of the mean of three journalists with a held-out journalist (0.57).}
    \label{fig-survey-equivalence-nyu-fringe-sites}
\end{figure*}
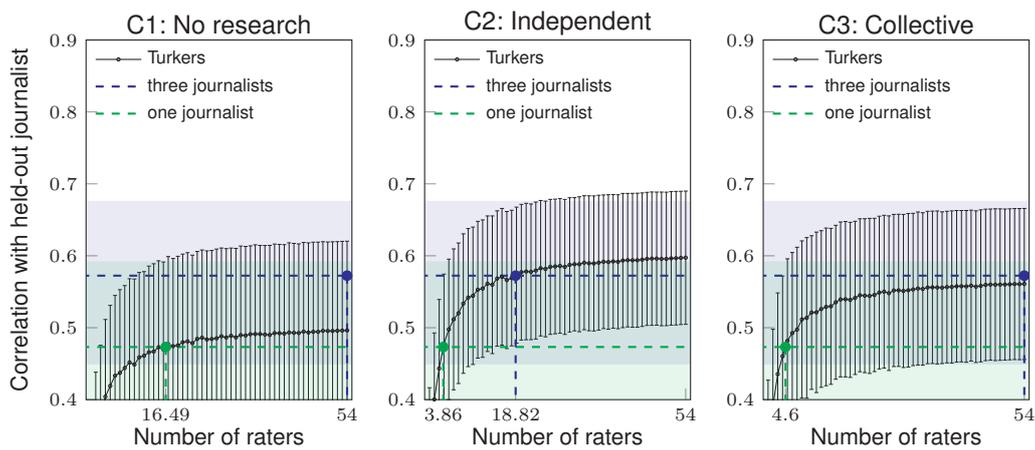

\FloatBarrier

\begin{table*}
\centering
\small
\begin{tabular}{lrrr}
 & \textbf{C1: No Research} & \textbf{C2: Individual Research} & \textbf{C3: Collective Research}  \\ [1ex]
 \midrule
Liberal-Conservative Correlation & .69 & .81 & .79  \\ [1ex] 
Correlation of one lay rater with one journalist  & 0.42  & 0.43 & 0.42 \\ [1ex] 
Lay raters equivalent to one journalist & 5.77  & 5.19 & 7.22 \\ [1ex] 
Lay raters equivalent to two journalists & 14.83  & 15.26 & $>$54 \\ [1ex] 
Liberal raters equivalent to one journalist & 2.20  & 2.61 & 2.69 \\ [1ex] 
Liberal raters equivalent to two journalists & 4.69  & 5.31 & 7.69 \\ [1ex] 
Conservative raters equivalent to one journalist & $>$18  & $>$18 & $>$18 \\ [1ex] 
Conservative raters equivalent to two journalists & $>$18  & $>$18 & $>$18 \\ [1ex] 
\end{tabular}
\vspace{4pt}
\caption{Summary of results for only political articles from collection one, using our four journalist ratings.}
\label{tab:results-summary-political-items}
\end{table*}

\begin{figure*}
\centering
\includegraphics[width=0.8\textwidth]{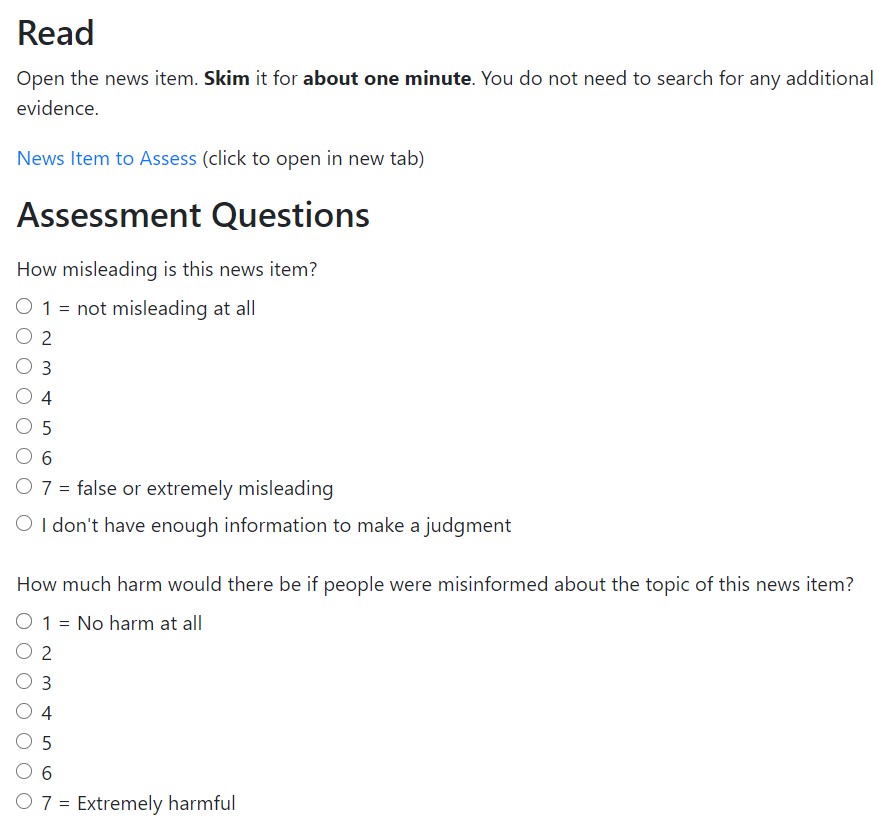}
\caption{Labeling interface for Condition 1: No Research.}
\label{fig:screenshot-C1}
\end{figure*}

\begin{figure*}
\centering
\includegraphics[width=0.8\textwidth]{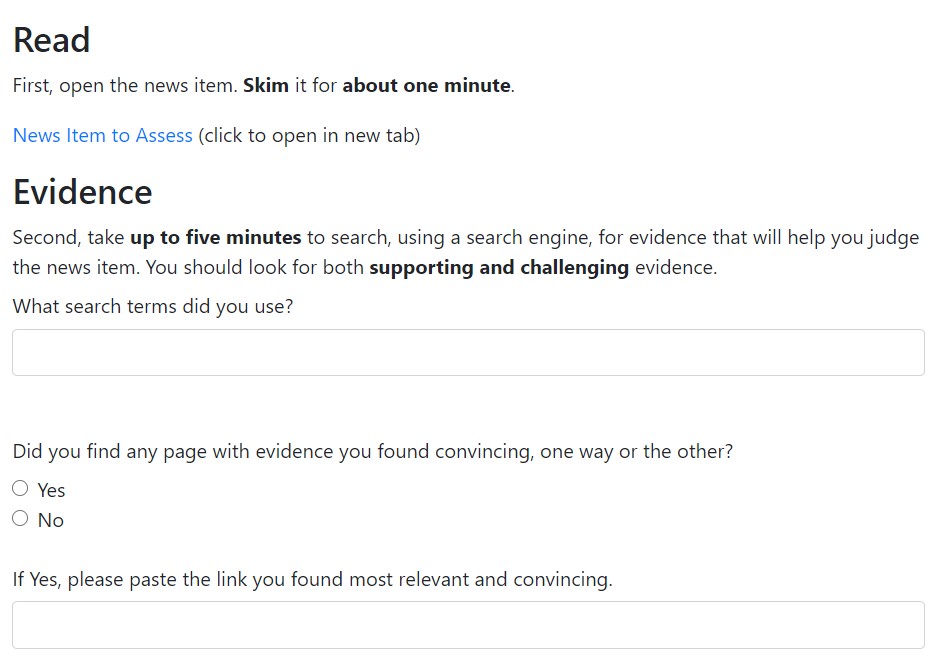}
\caption{Labeling interface for Condition 2: Independent Research.}
\label{fig:screenshot-C2}
\end{figure*}

\begin{figure*}
\centering
\includegraphics[width=0.8\textwidth]{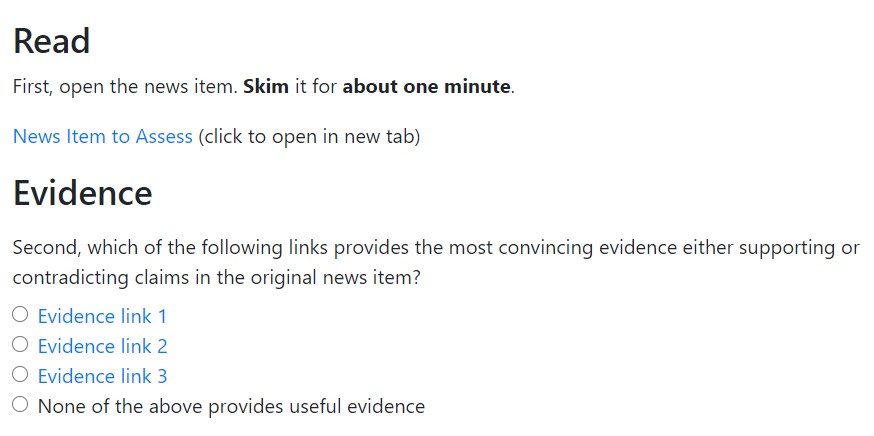}
\caption{Labeling interface for Condition 3: Collective Research.}
\label{fig:screenshot-C3}
\end{figure*}

\begin{figure*}
\centering
\includegraphics[width=0.6\textwidth]{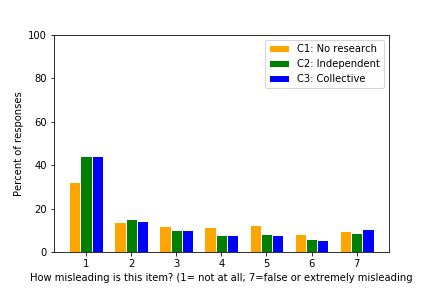}
\caption{Frequency of Judgments By Condition}
\label{fig:frequency-of-judgments-by-condition}
\end{figure*}

\begin{figure*}
\centering
\includegraphics[width=0.6\textwidth]{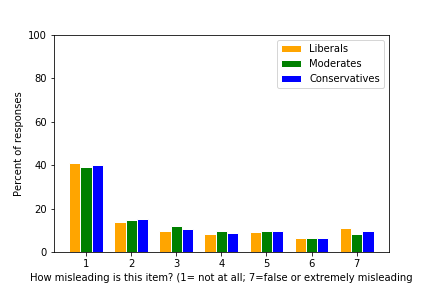}
\caption{Frequency of Judgments By Ideology}
\label{fig:frequency-of-judgments-by-ideology}
\end{figure*}

\newpage

\begin{figure*}
\centering
\includegraphics[width=0.6\textwidth]{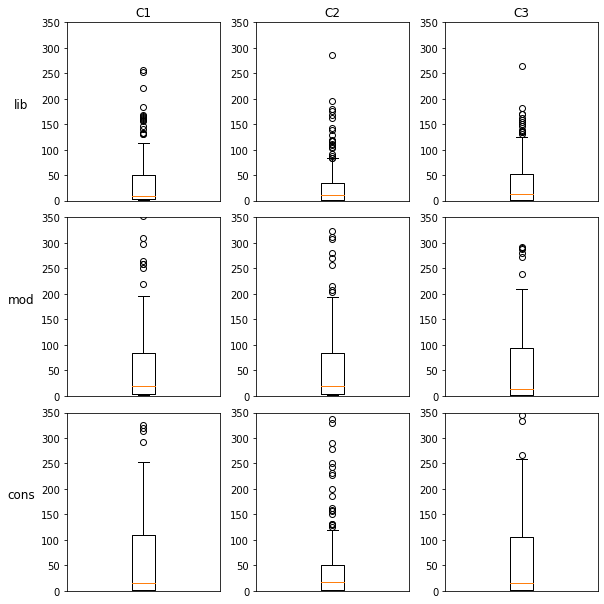}
\caption{Distribution of number of items rated by lay raters.}
\label{fig:item_rated_num}
\end{figure*}

\newpage

\FloatBarrier

\footnotesize
\clearpage
\onecolumn
\begin{longtable}{p{17cm}}
\textbf{URL} \\ \midrule
https://worldtruth.tv/wall-street-journal-investigation-finds-amazon-com-selling-dumpster-trash-food-supplements-as-new/ \\
https://neonnettle.com/news/4335-obama-at-bilderberg-the-us-must-surrender-to-the-new-world-order- \\
https://socialsecurityworks.org/2018/04/12/politicians-steal-social-security/ \\
https://www.foxnews.com/politics/trump-impeachment-defense-hunter-biden-john-bolton-senate \\
https://crooksandliars.com/2020/02/susan-collins-betrays-country-vote-acquit \\
https://www.cnbc.com/2020/01/21/cdc-to-announce-first-us-case-of-china-coronavirus-that-has-killed-6-cnn-reports.html?\_\_source=facebook\%7Cmain\\\&fbclid=IwAR1cKn8eEKKfKLqSqfYE2nK-5kZyDLGKSSWOhUndr99vl1-L2ecI7xHv3Dg\&fbclid=IwAR21zcbYAnqDt3nFWpYbZ1CZuUoq4-NxghFC\_9tlgmoPUYvjnK8sEorpKNQ \\
https://meaww.com/queen-invites-meghan-markle-mother-doria-christmas-sandringham \\
https://www.nytimes.com/2018/05/22/us/politics/georgia-primary-abrams-results.html \\
https://www.rd.com/food/fun/coffee-black-psychopath-study/ \\
https://www.palmerreport.com/analysis/the-sealed-indictment-a-that-donald-trump-needs-to-worry-about-more-than-ever/23359/ \\
http://www.msnbc.com/rachel-maddow-show/trump-crafts-new-excuse-massive-budget-deficit-forest-fires \\
https://www.nbcnews.com/news/us-news/17-kids-injured-after-delta-jet-dumps-fuel-l-playground-n1115586?cid=sm\_npd\_nn\_fb\_ma \\
https://www.collective-evolution.com/2020/01/22/another-supposedly-authentic-photo-of-a-ufo-the-story-behind-it/ \\
https://ijr.com/schiff-claims-does-not-know-whistleblower-impeachment-hearing/ \\
https://www.nytimes.com/2020/01/06/us/politics/trump-esper-iran-cultural-sites.html?smid=fb-nytimes\&smtyp=cur \\
https://www.washingtontimes.com/news/2020/jan/26/coronavirus-link-china-biowarfare-program-possible/ \\
https://bipartisanreport.com/2019/12/07/ex-intel-slam-trump-for-sucking-up-to-saudis-after-navy-shooting/ \\
https://babylonbee.com/news/bill-clinton-allegations-of-sexual-misconduct-should-disqualify-a-man-from-public-office \\
https://www.zerohedge.com/markets/china-says-us-must-reduce-tariffs-phase-one-trade-deal-amid-stalled-talks \\
https://www.bet.com/celebrities/news/2016/12/07/kim-kardashian-kanye-west-marriage.html?cid=facebook \\
https://www.huffpost.com/entry/kamala-harris-dropping-out\_n\_5de6a263e4b0913e6f8684bd?ncid=fcbklnkushpmg00000063 \\ \&utm\_source=main\_fb\&utm\_medium=facebook\&utm\_campaign=hp\_fb\_pages \\
http://www.ladbible.com/community/interesting-a-giant-500ft-asteroid-is-heading-for-earth-at-20000mph-20180826 \\
https://www.westernjournal.com/obama-urges-lives-better-non-americans/ \\
https://www.westernjournal.com/ct/trump-black-liberal-reporter-racist-question/ \\
https://www.cbsnews.com/news/iguana-florida-falling-weather-service-issues-alert-as-temperatures-drop-2020-01-21/ \\
http://www.social-consciousness.com/2017/03/donald-trump-says-pedophiles-will-get-death-penalty.html \\
https://themindunleashed.com/2020/01/death-tolls-rises-106-americans-evacuate-wuhan.html \\
https://www.teaparty.org/ukrainian-officials-release-records-of-46-payments-to-hunter-biden-from-burisma-holdings-38-payments-were-for-83333-totaling-over-3-1-million-416961/ \\
https://www.dcclothesline.com/2020/01/27/lawmakers-pushing-to-make-michigan-a-2nd-amendment-sanctuary-state/ \\
https://www.zerohedge.com/geopolitical/china-quarantines-second-city-experts-warn-its-already-too-late-stop-virus \\
https://www.yahoo.com/news/trump-helped-parents-hide-money-tax-returns-york-195327106.html \\
http://www.ladbible.com/entertainment/music-fans-flip-the-finger-at-machine-gun-kelly-and-get-in-his-face-20180920 \\
https://www.rollingstone.com/politics/politics-news/brian-kemp-leaked-audio-georgia-voting-745711/ \\
https://patriotjournal.org/democrat-fbi-closet/ \\
http://www.breakingnews247.net/5b872f00c5639/massive-alligator-found-in-browns-mills-new-jersey.html \\
https://www.breitbart.com/politics/2018/11/01/orourke-campaign-exposed-in-undercover-video-for-assisting-honduran-migrants-nobody-needs-to-know/ \\
https://www.healthnutnews.com/60-lab-studies-now-confirm-cancer-link-to-a-vaccine-you-probably-had-as-a-child/ \\
https://www.nbcnews.com/pop-culture/celebrity/burt-reynolds-charismatic-star-1970s-blockbusters-dies-82-n907206 \\
https://www.dailywire.com/news/35631/linda-sarsour-calls-dehumanization-jews-report-ryan-saavedra \\
https://fellowshipoftheminds.com/pentagon-bans-bible-verses-on-dog-tags-while-pres-trump-upholds-right-to-pray-in-public-schools \\
http://www.theguardian.com/us-news/2018/nov/10/trump-baltics-balkans-mixup-le-monde-belleau-cemetery-paris \\
http://coolcatapproves.com/funny/australia-doesnt-exist-and-people-who-live-there-are-actors-paid-by-nasa-flat-earthers-claim/ \\
https://www.dailymail.co.uk/news/article-6373429/Caitlyn-Jenners-Malibu-house-burns-California-wildfires-rage-control.html \\
https://www.usatoday.com/story/news/politics/2018/08/30/federal-pay-freeze-trump-cancels-2-1-percent-pay-raise-federal-workers-citing-budget-deficit/1145355002/ \\
https://www.huffpost.com/entry/art-acevedo-ted-cruz-mitch-mcconnell-nra-gun-victims\_n\_5def2944e4b05d1e8a56fa2f?section=politics \\ \&utm\_campaign=hp\_fb\_pages\&utm\_source=main\_fb\&utm\_medium=facebook\&ncid=fcbklnkushpmg00000063 \\
https://www.cnbc.com/2018/10/19/saudi-arabia-admits-journalist-jamal-khashoggi-was-killed-after-a-fight-broke-out-in-consulate.html \\
https://news.unclesamsmisguidedchildren.com/obama-says-benghazi-is-a-wild-conspiracy-theory/ \\
https://worldnewsdailyreport.com/cops-beat-up-teen-after-bank-teller-mistakes-his-erection-for-a-pistol/ \\
https://clashdaily.com/2019/12/ukraine-advisor-disputes-key-point-in-impeachment-testimony-is-this-bad-news-for-democrats/ \\
https://ijr.com/graham-eyes-rule-change-kick-start-trump-trial/ \\
https://bipartisanreport.com/2020/01/12/new-trump-approval-poll-released-confirms-massive-2020-blue-wave/ \\
https://themindunleashed.com/2020/01/300-vultures-occupy-border-patrol-tower-covering-it-with-corrosive-feces-vomit.html \\
http://www.ladbible.com/entertainment/uk-boyfriend-runs-away-after-his-girlfriend-catches-a-brides-bouquet-20180904 \\
https://www.breitbart.com/big-journalism/2018/08/18/cnn-accused-intimidating-paul-manafort-jury/ \\
https://www.nytimes.com/2019/12/04/world/asia/afghanistan-tetsu-nakamura-dead.html?smid=fb-nytimes\&smtyp=cur\&fbclid=\\IwAR3fKvdPsRQ4wUNylstquIFOOj7zC64WmA4eXrrNp6XnSUJIk79oQ\_uSKqA\\\&fbclid=IwAR1iI65LDkftMGynOURUC\_ope2fTdXEAfdsRWnufyMmVSs7JN6Hcy6u86t8\\\&fbclid=IwAR2oZkicTvpbHfqtaDjl9QdZ-PQpy4FqZUwYpiky-5t26i21Kk7i7O6McUw \\
https://wokesloth.com/fox-news-poll/stefanie/ \\
https://www.vice.com/en\_us/article/43ejmj/generation-z-is-skipping-college-for-trade-school \\
https://www.cbsnews.com/news/kanye-west-joel-osteen-lakewood-church-kim-kardashian-church-service-nebuchadnezzar/?ftag=CNM-00-10aab6a\&linkId=77264473 \\
https://www.express.co.uk/news/world/1232244/coronavirus-outbreak-china-Huanggang-Wuhan-coronavirus-spread-latest \\
https://www.nowtheendbegins.com/kanye-west-sunday-service-jesus-is-king-lakewood-church-all-seeing-eye-horus-illuminati/ \\
https://www.theroot.com/trump-rants-like-racist-grandpa-in-speech-to-maga-negro-1830045693 \\
https://www.washingtonpost.com/sports/2019/12/09/megan-rapinoe-is-sports-illustrateds-sportsperson-year-only-fourth-woman-chosen-alone/?tid=sm\_fb \\
https://www.usmagazine.com/celebrity-news/news/burt-reynolds-dead-actor-dies-at-82/ \\
https://www.cnbc.com/2019/12/18/trump-impeached-by-house-for-high-crimes-and-misdemeanors.html?\_\_source=facebook\%7Cmain \\
https://news.vice.com/en\_us/article/3kek75/another-kavanaugh-accuser-is-taking-to-maryland-authorities \\
https://worldnewsdailyreport.com/woman-arrested-for-training-squirrels-to-attack-her-ex-boyfriend/ \\
https://www.dailywire.com/news/35040/controversial-reporter-jemele-hill-out-espn-part-emily-zanotti \\
https://legalinsurrection.com/2018/09/maxine-waters-suggests-knocking-off-trump-then-going-after-pence/ \\
https://www.newyorker.com/news/news-desk/the-five-year-old-who-was-detained-at-the-border-and-convinced-to-sign-away-her-rights \\
https://themindunleashed.com/2019/11/family-facing-jail-living-rv-their-own-property-repair-home-after-fire.html \\
https://www.thedailybeast.com/ivanka-trumps-gurus-say-their-techniques-can-end-war-and-make-you-fly \\
https://www.iflscience.com/plants-and-animals/drone-footage-reveals-over-100-whales-trapped-in-secret-underwater-jails/ \\
https://www.washingtontimes.com/news/2020/jan/13/donald-trump-melania-trump-cheered-national-champi/ \\
https://www.thedenverchannel.com/news/chris-watts-case/chris-watts-reaches-plea-deal-to-avoid-death-penalty-in-deaths-of-pregnant-wife-2-daughters \\
https://www.teaparty.org/holy-moses-more-than-175000-tickets-requested-to-see-president-trump-in-new-jersey-supporters-line-up-48-hours-426092/ \\
https://www.zerohedge.com/geopolitical/russias-only-aircraft-carrier-admiral-kuznetsov-has-erupted-flames \\
https://www.yahoo.com/entertainment/orange-new-black-cancelled-netflix-085607387.html \\
https://www.dcclothesline.com/2019/12/04/in-2018-86-of-those-arrested-for-violent-crime-in-los-angeles-were-non-white-5-were-white-the-city-is-28-white/ \\
https://www.politico.com/news/2020/01/08/ruth-bader-ginsburg-health-096302 \\
https://www.nbcnews.com/video/cincinnati-police-tased-11-year-old-girl-accused-of-shoplifting-1313262659520?v=raila\& \\
https://www.zerohedge.com/markets/boeing-refused-request-more-737-max-pilot-training-deadly-crash \\
http://www.higherperspectives.com/one-glass-red-wine-1577145867.html \\
https://abcnews.go.com/Politics/desantis-floridians-monkey-electing-african-american-democrat-governor/story?id=57476957 \\
https://www.breitbart.com/politics/2018/10/14/cnn-poll-biden-leads-field-2020-democratic-hopefuls/ \\
https://lawandcrime.com/civil-rights/gop-removes-sole-polling-place-from-famous-hispanic-majority-city-in-kansas/ \\
https://www.nbcnews.com/politics/donald-trump/least-impressive-sex-i-ever-had-stormy-daniels-tells-all-n910566 \\
https://www.palmerreport.com/analysis/nancy-pelosi-knows-something-we-dont/23491/ \\
http://www.55meals.com/did-you-know-your-energy-drinks-contain-bull-urine-semen/ \\
https://www.washingtonpost.com/technology/2018/11/08/white-house-shares-doctored-video-support-punishment-journalist-jim-acosta/ \\
https://conservativedailypost.com/savage-claims-ford-deeply-tied-to-deep-state/ \\
https://conservativepost.com/these-democrats-voted-against-impeaching-trump/ \\
https://www.thegatewaypundit.com/2019/12/trans-activists-target-olympic-cyclist-inga-thompson-for-saying-women-shouldnt-have-to-compete-with-biological-men/ \\
https://www.dailywire.com/news/37685/epa-greenhouse-gas-emissions-dropped-nearly-3-joseph-curl \\
https://www.nytimes.com/2019/11/20/world/africa/almaas-elman-somalia.html?smtyp=cur\&smid=fb-nytimes \\
http://rare.us/rare-news/around-the-world/kids-cant-tell-time/ \\
https://www.cnn.com/2018/10/24/us/florida-middle-girls-allegedly-wanted-to-kill-classmates/index.html \\
https://www.cbsnews.com/news/serena-williams-tennis-wins-first-title-in-3-years-donates-prize-money-australia-wildfire-relief-2020-01-12/?ftag=CNM-00-10aab6a\&linkId=80518649 \\
https://ijr.com/pelosi-house-will-vote-resolution-limit-trumps-military-actions-iran/ \\
https://www.thisisinsider.com/michael-buble-quits-music-sons-cancer-battle-2018-10 \\
https://www.lovebscott.com/steve-harveys-talk-show-canceled-nbc-will-replaced-kelly-clarkson-show \\
https://www.foxnews.com/entertainment/cardi-b-trump-nigerian-citizenship \\
https://nypost.com/2019/12/08/walmart-apologizes-for-sweater-featuring-santa-with-cocaine/?utm\_medium=SocialFlow \\ \&sr\_share=facebook\&utm\_source=NYPFacebook\&utm\_campaign=SocialFlow \\
https://crooksandliars.com/2020/02/jennifer-granholm-catches-rick-santorum \\
https://www.newsweek.com/witches-hex-brett-kavanaugh-amy-kremer-scary-conservatives-ritual-trump-1168948 \\
https://www.nbcnews.com/video/sessions-says-he-will-recuse-himself-from-any-clinton-investigations-851626564002?v=railb \\
https://www.healthnutnews.com/nbc-abc-and-cbs-appear-to-have-run-cover-for-worlds-most-powerful-rape-rings/ \\
https://www.nbcnews.com/news/us-news/trump-tweets-all-well-after-iranian-missile-attack-targeting-u-n1112211?cid=sm\_npd\_nn\_fb\_ma\\\&fbclid=IwAR1PQV1l4jbUJUHS91zCB228EH97Bn9\_TSqMIQ9mRbFhvOVNNbL7WcYWsHg\\\&fbclid=IwAR2n15afcp34Lqppj6ssx7ysDyUhqDkjygYxOmNPH4b4oZ9RlWYGkPmSVqM\\\&fbclid=IwAR1PQV1l4jbUJUHS91zCB228EH97Bn9\_TSqMIQ9mRbFhvOVNNbL7WcYWsHg \\
https://ijr.com/schiff-vote-impeach-obama-engaged-similar-conduct/ \\
https://www.jta.org/2018/10/27/top-headlines/trump-blames-deaths-pittsburgh-synagogue-lack-armed-guards \\
http://time.com/5390884/nike-sales-go-up-kaepernick-ad/ \\
https://www.dailywire.com/news/35097/la-and-ny-overrun-topless-women-seeking-equality-amanda-prestigiacomo \\
https://www.cnbc.com/2020/01/03/bill-gates-americas-tax-system-is-not-fair.html?\_\_source=facebook\%7Cmain \\
https://townhall.com/tipsheet/juliorosas/2020/01/14/president-trump-gets-thunderous-applause-at-clemson-and-lsu-playoff-game-n2559475?utm\_content=bufferfe3f8\&utm\_medium=social\&utm\_source=facebook.com\&utm\_campaign=buffer \\
https://nypost.com/2020/02/05/trump-acquitted-in-senate-impeachment-trial/ \\ ?utm\_source=NYPFacebook\&utm\_medium=Native\&utm\_campaign=NYPFacebook \\
https://worldnewsdailyreport.com/man-accused-of-raping-a-cow-claims-it-is-the-reincarnation-of-his-dead-wife/ \\
https://barenakedislam.com/2019/12/05/ding-ding-ding-first-muslim-woman-elected-to-pennsylvania-house-of-representatives-has-been-arrested-for-stealing-500000-from-a-charity/ \\
https://patriotjournal.org/sanctuary-state-vote-trump/ \\
http://alexschadenberg.blogspot.com/2018/10/sick-kids-hospital-toronto-will.html \\
https://bipartisanreport.com/2020/01/06/schiff-hammers-president-gop-over-impeachment-trial-obstruction/ \\
https://conservativefiringline.com/video-of-the-day-dem-rep-raskin-thanks-congressman-helping-form-rules-for-sham-impeachment-of-trump-who-was-impeached-for-bribery/ \\
https://news.vice.com/en\_us/article/nepwng/some-texans-had-to-wait-so-long-to-vote-they-gave-up-a-lawsuit-is-trying-to-give-them-a-second-chance \\
https://tribunist.com/news/active-shooter-commits-suicide-after-being-confronted-by-armed-citizen-at-ok-walmart/ \\
https://www.hannity.com/media-room/free-money-cory-booker-unveils-plan-to-give-poor-americans-50000/ \\
https://crooksandliars.com/2020/01/john-bolton-will-testify-if-subpoenaed-so \\
https://realfarmacy.com/surgeon-mammogram/ \\
https://www.usatoday.com/story/news/nation-now/2018/08/18/no-prison-time-ex-houston-doctor-who-raped-heavily-sedated-patient/1031665002/ \\
https://abcnews.go.com/Politics/blame-abc-news-finds-17-cases-invoking-trump/story?id=58912889 \\
https://www.foxnews.com/us/virginia-capitol-gun-rights-rally \\
https://townhall.com/tipsheet/juliorosas/2019/12/09/joe-biden-claims-no-one-told-him-about-potential-conflict-of-interest-with-hunters-job-at-burisma-n2557688/?utm\_content=buffer644e4\&utm\_medium=social\&utm\_source=facebook.com\&utm\_campaign=buffer \\
http://www.tmz.com/2018/09/18/sesame-street-bert-ernie-gay-couple-confirmed-writer-speculation-over/ \\
https://www.thedailybeast.com/elizabeth-warren-releases-native-american-dna-test-results \\
https://nypost.com/2020/01/23/three-us-firefighters-killed-in-australia-plane-crash-while-battling-wildfires/?utm\_source=NYPFacebook\&sr\_share=facebook\\\&utm\_medium=SocialFlow\&utm\_campaign=SocialFlow \\
https://www.theguardian.com/environment/2019/nov/13/waves-in-st-marks-square-as-venice-flooded-highest-tide-in-50-years?CMP=fb\_gu\&utm\_medium=Social\&utm\_source=Facebook\#Echobox=1573628711 \\
https://www.dcclothesline.com/2020/01/14/wisconsin-judge-orders-up-to-209000-listings-purged-from-voter-rolls-finds-3-in-contempt-orders-fines-for-delay/ \\
https://www.aol.com/article/news/2015/12/09/study-links-e-cigarettes-to-incurable-disease-called-popcorn-lu/21281029/ \\
https://www.lifenews.com/2018/08/28/united-methodist-church-proposes-new-position-statement-saying-we-support-abortion/ \\
http://nypost.com/2018/10/09/de-blasio-signs-bill-allowing-third-gender-on-birth-certificates/ \\
https://www.daily-sun.com/post/453807/Alaska-man-survives-three-weeks-with-little-food-and-shelter \\
https://www.theguardian.com/us-news/2019/dec/11/donald-trump-jr-mongolia-rare-endangered-sheep-permit?CMP=fb\_gu\&utm\_medium=Social\\\&utm\_source=Facebook\#Echobox=1576106294 \\
https://clashdaily.com/2019/11/schiff-named-in-wh-officials-defamation-lawsuit-leaked-lies-to-politico-to-push-impeachment/ \\
https://www.iflscience.com/health-and-medicine/a-cancer-kill-switch-has-been-found-in-the-body-and-researchers-are-already-hard-at-work-to-harness-it/ \\
https://www.palmerreport.com/news/no-win-cornr-gop-senators-donald-trump/24395/ \\
https://educateinspirechange.org/health/experienced-butcher-admits-see-cancer-pork-just-cut-still-sell-customers/ \\
https://americanmilitarynews.com/2018/10/armys-nco-promotion-criteria-fails-to-gauge-leadership-qualities-study-says/ \\
https://www.dailywire.com/news/38230/broward-county-elections-supervisor-be-forced-james-barrett \\
https://kutv.com/news/local/woman-who-alleges-mtc-president-raped-her-filmed-testifying-about-rape-in-church \\
https://www.yahoo.com/news/barack-obama-women-singapore-115412140.html?ncid=facebook\_yahoonewsf\_akfmevaatca \\
https://www.newsandguts.com/white-house-leaves-pence-condemn-attacks-clinton-obama-cnn/ \\
https://crooksandliars.com/2020/01/joni-ernst-gives-away-ballgame-joe-biden \\
https://www.zerohedge.com/geopolitical/ukrainian-indictment-reveals-hunter-biden-group-made-165-million-mp \\
https://abcnews.go.com/Politics/trump-kicks-off-week-tweet-calling-media-true/story?id=58827743 \\
https://www.nbcnews.com/news/us-news/retired-firefighter-who-fired-shotgun-black-teen-asking-directions-gets-n935611 \\
https://www.thenewamerican.com/usnews/crime/item/24549-nypd-source-weiner-laptop-has-enough-evidence-to-put-hillary-away-for-life \\
https://www.dailywire.com/news/36198/watch-beto-orourke-pins-himself-corner-over-drunk-ryan-saavedra \\
https://www.naturalnews.com/2020-01-20-san-fran-democrat-tyrants-taxing-landlords-leaving-stores-vacant.html \\
https://crooksandliars.com/2020/01/even-c-span-ticked-over-impeachment \\
https://www.breitbart.com/midterm-election/2018/10/18/nancy-pelosi-collateral-damage/ \\
https://www.christianpost.com/news/white-house-hosts-100-evangelical-leaders-state-like-dinner-this-is-spiritual-warfare-227044/ \\
https://www.collective-evolution.com/2020/01/09/the-us-military-pollutes-more-140-countries-combined/ \\
https://bipartisanreport.com/2019/11/18/doctor-tells-cnn-trumps-walter-reed-medical-visit-was-fishy/ \\
https://www.theepochtimes.com/joy-disbelief-as-korean-families-separated-by-war-meet-after-65-years\_2628430.html \\
https://nypost.com/2019/11/29/f-k-white-people-graffiti-found-outside-queens-home/?sr\_share=facebook\&utm\_medium=SocialFlow \\ \&utm\_source=NYPFacebook\&utm\_campaign=SocialFlow \\
https://tribunist.com/news/band-performs-skit-about-shooting-police-during-halftime-of-football-game/ \\
https://abc13.com/4038817/ \\
https://www.palmerreport.com/analysis/everything-is-falling-apart-for-donald-trump-in-real-time/24192/ \\
https://occupydemocrats.com/2019/12/10/bushs-ethics-chief-trumps-are-an-organized-crime-family-we-need-to-go-after-all-of-them/ \\
https://www.dailywire.com/news/37552/migrant-caravan-marching-us-borders-swells-14000-joseph-curl \\
https://occupydemocrats.com/2020/01/14/newly-released-texts-from-giuliani-crony-parnas-appear-to-show-them-spying-on-amb-yovanovich-4/ \\
https://www.lifenews.com/2018/08/27/125-women-take-abortion-pills-to-kill-their-babies-to-protest-pro-life-laws/ \\
https://thefederalistpapers.org/opinion/california-bans-feeding-homeless-without-police-supervision \\
https://www.lapd.com/blog/lapd-officer-shot-point-blank-range-and-not-peep-aclu-blm-or-assemblymember-weber \\
https://ijr.com/trump-poised-week-become-third-president-impeached/ \\
https://www.pbs.org/newshour/politics/trump-says-it-will-be-hard-to-unify-country-without-a-major-event \\
https://worldnewsdailyreport.com/morgue-worker-arrested-after-giving-birth-to-a-dead-mans-baby/ \\
https://www.huffingtonpost.com.au/entry/woman-saves-scorched-koala-with-shirt\_au\_5dd474aae4b0fc53f20a6a44 \\ ?ncid=fcbklnkushpmg00000063\&utm\_campaign=hp\_fb\_pages\&utm\_medium=facebook\&utm\_source=main\_fb \\
https://www.myleaderpaper.com/news/accidents/desloge-man-dies-in-two-vehicle-accident-west-of-festus/article\_f2b6b228-dbdd-11e8-8b87-abb2d7153695.html \\
https://www.washingtontimes.com/news/2018/oct/14/ted-wheeler-portland-mayor-stands-decision-allow-a/ \\
https://www.lifenews.com/2018/09/14/chelsea-clinton-says-it-would-be-un-christian-to-protect-babies-from-abortion/ \\
https://ijr.com/smollet-distress-malicious-prosecution-lawsuit-chicago/?utm\_campaign=IJR\%20-\%20Facebook\%20Content\&utm\_content=106041134\\\&utm\_medium=social\&utm\_source=facebook\&hss\_channel=fbp-189885532970 \\
https://www.thehollywoodgossip.com/2018/07/meghan-markle-and-kate-middleton-due-to-give-birth-on-the-same-d/ \\
https://www.teaparty.org/breaking-ukrainian-official-reveals-six-criminal-cases-opened-in-ukraine-involving-the-bidens-420208/ \\
https://www.teaparty.org/back-home-in-pelosis-san-francisco-homeless-drug-addicts-are-now-taking-dumps-in-the-supermarket-aisles-421170/ \\
https://www.msn.com/en-us/news/us/jewish-leaders-tell-trump-hes-not-welcome-in-pittsburgh-until-he-denounces-white-nationalism/ar-BBP2lEL \\
https://www.nbcnews.com/news/us-news/meijer-pharmacist-denies-michigan-woman-miscarriage-medication-citing-religious-beliefs-n921711 \\
https://www.palmerreport.com/analysis/trumps-sham-acquittal-is-already-blowing-up-in-senate-republicans-faces/24893/ \\
https://prepforthat.com/christine-blasey-fords-yearbook-seems-to-show-high-school-racism/ \\
https://www.express.co.uk/news/uk/1212362/Donald-Trump-News-Jeremy-Corbyn-General-Election-Boris-Johnson-NHS \\
https://www.breitbart.com/video/2018/10/18/biden-there-is-no-evidence-of-widespread-voter-fraud-in-the-american-electoral-process/ \\
https://www.infowars.com/leaked-french-internal-intelligence-report-claims-150-neighborhoods-held-by-radical-islamists/ \\
https://www.infowars.com/never-trumper-rick-wilson-suggests-putting-anti-vaxxers-in-re-education-camps/ \\
https://occupydemocrats.com/2019/11/20/sondlands-testimony-directly-implicates-trump-pence-and-pompeo-in-ukraine-quid-pro-quo-plot/ \\
https://www.breitbart.com/video/2018/10/22/obama-unlike-some-i-state-facts-i-dont-believe-in-just-making-stuff-up/ \\
https://abcnews.go.com/Politics/trump-plans-end-birthright-citizenship-babies-born-citizens/story?id=58845684 \\
https://crooksandliars.com/2019/11/roger-stone-trial-ends-rick-gates \\
https://www.nbcnews.com/news/crime-courts/nine-shot-multiple-killed-family-gathering-fresno-california-n1084706?cid=sm\_npd\_nn\_fb\_ma \\
https://www.teaparty.org/indictment-against-head-of-burisma-reveals-hunter-biden-was-receiving-payments-from-money-raised-through-criminal-means-siphoned-laundered-from-ukraine-417981/ \\
https://www.yahoo.com/news/duncan-hunter-guilty-plea-190447228.html?ncid=facebook\_yahoonewsf\_akfmevaatca \\
https://www.infowars.com/nearly-200-people-arrested-across-australia-for-deliberately-starting-bushfires/ \\
https://bipartisanreport.com/2019/12/01/senator-announces-impeach-vote-hint-that-has-trump-fuming/ \\
https://www.huffpost.com/entry/katie-sowers-first-female-gay-coach-super-bowl\_n\_5e25ebf5c5b632117617c0d1 \\ ?utm\_campaign=hp\_fb\_pages\&utm\_source=main\_fb\&ncid=fcbklnkushpmg00000063\&utm\_medium=facebook\&ir= \\ Entertainment\&guccounter=1\&guce\_referrer=aHR0cHM6Ly9hcHBzLmNyb3dkdGFuZ2xlLmNvbS9zbWFwcG55dS9saXN0cy8xMzAyMjc1 \\ \&guce\_referrer\_sig=AQAAAHlDY7JuKLcaFb1J7U2k6To8Eb6WqEIcvF2UYk7n18QxQFLNlTsBlrwmkEulOWENXn6vFnEofr4KTyVO6oYXD \\ sf8HM0X8CRIkIh35zaSFPCtWQVO8rJc7tfSAzPvT0Q0XAJ6HewH8l7avdgCZRa1IAw\_caZeJuKPdZ0Ct-kgKpTd \\
https://www.thegatewaypundit.com/2018/09/former-scalia-law-clerk-drops-pictures-and-evidence-that-blows-christine-fords-case-wide-open/ \\
https://www.iflscience.com/editors-blog/pieces-of-a-ufo-fell-from-the-sky-and-landed-in-remote-cambodian-village/ \\
http://www.tmz.com/2018/11/01/president-trump-immigrant-caravan-throwing-rocks-same-firearms/ \\
https://www.cnbc.com/2019/12/10/house-democrats-announce-articles-of-impeachment-against-trump.html?\_\_source=facebook\%7Cmain \\
https://www.newsandguts.com/white-house-uses-doctored-video-support-claim-reporter/ \\
http://nypost.com/2018/09/27/two-men-tell-senate-that-they-not-kavanaugh-assaulted-ford/ \\
https://stillnessinthestorm.com/2020/01/key-brain-region-smaller-in-birth-control-pill-users/ \\
http://endoftheamericandream.com/archives/why-are-volcanoes-all-over-the-globe-suddenly-shooting-giant-clouds-of-ash-miles-into-the-air \\
https://www.cbsnews.com/live-updates/trump-impeachment-vote-live-stream-today-house-of-representatives-articles-of-impeachment-2019-12-18/ \\
https://www.concealedcarry.com/news/armed-citizens-are-successful-95-of-the-time-at-active-shooter-events-fbi/ \\
https://www.nbcnews.com/pop-culture/pop-culture-news/caroll-spinney-legendary-sesame-street-puppeteer-big-bird-dies-85-n1097806?cid=sm\_npd\_nn\_fb\_ma \\
https://www.westernjournal.com/ct/huge-feinstein-investigated/ \\
https://www.christianpost.com/news/china-trying-to-rewrite-the-bible-force-churches-sing-communist-anthems-227664/ \\
https://www.zerohedge.com/geopolitical/iran-evaluating-13-retaliation-scenarios-inflict-historic-nightmare-us \\
https://shareblue.com/russia-president-vladimir-putin-annual-address-americas-downfall/ \\
https://bipartisanreport.com/2020/01/21/schiff-opening-impeachment-trial-statement-to-go-down-in-history/ \\
https://www.breitbart.com/the-media/2018/10/26/nolte-nbc-news-hid-info-wouldve-cleared-kavanaugh-avenatti-rape-allegations/ \\
https://www.pbs.org/newshour/politics/watch-kanye-west-in-white-house-visit-says-maga-hat-gives-him-power \\
https://www.foxnews.com/politics/mob-chants-threats-outside-tucker-carlsons-dc-home \\
https://www.foxnews.com/media/former-nfl-player-marine-jeremy-staat-kaepernick-disgrace \\
https://www.wthr.com/article/pox-parties-latest-rage-parents-trying-immunize-kids-against-chickenpox \\
https://thefreethoughtproject.com/watch-cop-chokes-innocent-man-calling-water-company-corruption/ \\
https://www.palmerreport.com/analysis/retweeting-bizarre-fake-account/23745/ \\
https://occupydemocrats.com/2020/01/26/damning-potential-john-bolton-ukraine-impeachment-testimony-revealed-in-early-leak-of-book-draft/ \\
https://www.politicususa.com/2018/10/26/texas-voting-machines-beto-orourke-ted-cruz.html \\
https://clashdaily.com/2020/02/john-kerry-says-that-the-entire-obama-admin-was-trying-to-get-rid-of-the-burisma-prosecutor/ \\
https://www.joyscribe.com/all-of-the-harry-potter-films-are-officially-coming-to-netflix/ \\
https://www.billboard.com/articles/columns/chart-beat/8094819/cardi-b-beyonce-five-hits-top-10-rb-hip-hop-songs-chart \\
https://www.cbsnews.com/live-updates/state-of-the-union-2020-president-donald-trump-tonight-2020-02-04-live-streaming/?ftag=CNM-00-10aab6a\&linkId=81729456 \\
http://www.collective-evolution.com/2017/02/13/how-sunscreen-could-be-causing-skin-cancer-not-the-sun/ \\
https://www.buzzfeed.com/aliciabarron/this-reeses-machine-will-swap-out-your-halloween-candy-for \\
https://occupydemocrats.com/2019/12/04/trump-caught-by-reporters-patting-himself-on-back-for-insulting-justin-trudeau/ \\
https://crooksandliars.com/2019/12/tucker-carlsons-white-power-hour-guest \\
https://www.nbcnews.com/news/us-news/video-shows-girl-16-apparently-kidnapped-nyc-street-front-mom-n1103176?cid=sm\_npd\_nn\_fb\_ma \\
https://www.yahoo.com/news/kobe-bryants-daughter-gianna-13-205036060.html \\
https://twitter.com/realDonaldTrump/status/1040217897703026689 \\
https://www.scarymommy.com/parkland-shooting-victim-sculpture/ \\
https://www.breitbart.com/border/2018/10/30/armed-migrants-in-caravan-opened-fire-on-mexican-cops-say-authorities/ \\
https://www.cnn.com/2018/10/16/us/first-female-commander-us-army-trnd/index.html \\
https://friendsforsyria.com/2019/12/04/ukrainian-neo-nazis-help-out-at-hong-kong-riots-pan-democrats-defend-them/ \\
https://www.theguardian.com/world/2020/jan/22/holocaust-survey-americans-pew-research-center?CMP=fb\_gu\&utm\_medium=Social \\ \&utm\_source=Facebook\#Echobox=1579708533 \\
https://www.newsweek.com/trump-ceo-pay-wages-tax-cuts-1076795 \\
https://www.cnbc.com/2019/08/05/hyundai-launches-car-with-a-roof-based-solar-charging-system.html?\_\_source=facebook\%7Cmain \\ \&fbclid=IwAR1HXryUvlbaPDWs-J39LnS2iksoD4qAO0mdai64d3znOCymsklN45IeUuY\&fbclid=IwAR3qBgscvE4tvqjZrDAW\_WdK-JWuBGDrjATWQgFYSQhepxMIp1x8PQK6p\_c\&fbclid=IwAR3Y3pIr8xjE7Ig7Wpz56Is\_c4Td538NWF80KyZlchePTjKOAn9m6T6qZ6I \\
https://www.dailywire.com/news/38153/breaking-voter-fraud-allegedly-found-deep-blue-ryan-saavedra \\
https://dailycaller.com/2018/10/15/elizabeth-warren-less-native-american-dna/ \\
https://www.movieguide.org/news-articles/netflix-animated-series-dedicates-an-entire-episode-to-promote-planned-parenthood-and-killing-babies.html \\
https://www.collective-evolution.com/2019/11/29/bestselling-novelist-who-wrote-about-vaccine-industry-deception-found-dead/ \\
https://ijr.com/democratic-white-house-race-messy-start-iowa/?utm\_campaign=IJR\%20-\%20Facebook\%20Content \\ \&utm\_content=115077734\&utm\_medium=social\&utm\_source=facebook\&hss\_channel=fbp-189885532970 \\
https://www.theepochtimes.com/threat-of-yellowstone-volcano-eruption-increases-to-high-says-usgs\_2700008.html \\
https://www.nbcnews.com/politics/trump-impeachment-inquiry/white-house-will-not-participate-wednesday-s-impeachment-hearing-n1093821?cid=sm\_npd\_nn\_fb\_ma \\
https://www.express.co.uk/news/world/1224973/iran-stampede-funeral-Qassem-Soleimani-pictures-latest-world-war-3 \\
https://news.jamaicans.com/kanye-west-bob-marleys-spirit-flows/ \\
https://www.naturalnews.com/2020-02-05-lies-we-are-being-told-about-coronavirus.html \\
https://www.eutimes.net/2019/11/donald-trump-jr-humiliated-booed-off-stage-by-angry-maga-voters/ \\
https://rewire.news/ablc/2018/10/11/supreme-court-native-americans-november/ \\
https://tribunist.com/news/americans-bought-enough-guns-on-black-friday-to-arm-the-marine-corps-yet-again/ \\
https://thefederalistpapers.org/us/liberal-professor-says-hurricane-victims-deserved-got-supporting-republicans \\
https://www.dailymail.co.uk/news/article-6285989/Canada-kicks-muted-pot-party-1st-G7-nation-OK-recreational-cannabis.html \\
https://themindunleashed.com/2020/02/earth-is-about-to-enter-a-30-year-mini-ice-age-as-the-sun-hibernates-scientist-warns.html \\
https://worldtruth.tv/the-coronavirus-was-engineered-by-scientists-in-a-lab-using-well-documented-genetic-engineering-vectors-that-leave-behind-a-fingerprint/ \\
https://www.lifezette.com/2018/11/trump-challenges-dems-on-kavanaugh-accusers-admission-that-she-made-it-all-up-where-are-you-on-this/ \\
http://www.latimes.com/local/lanow/la-me-ln-thousand-oaks-20181107-story.html \\
https://americanmilitarynews.com/2018/10/guatemala-captured-100-isis-terrorists-president-reveals-ahead-of-migrant-caravan-arrival/ \\
https://abc13.com/4554120/ \\
https://www.forbes.com/sites/rachellebergstein/2016/11/10/new-balance-gets-political-and-throws-support-behind-trump/ \\
http://fortune.com/2016/09/01/medical-marijuana-gun/ \\
https://worldtruth.tv/nature-science-journal-warned-about-pathogens-escaping-wuhan-level-4-biosafety-lab-bsl-4-before-coronavirus-outbreak/ \\
https://www.hatetriots.com/2018/10/nancy-pelosi-shouted-out-of-restaurant.html \\
https://www.cnbc.com/2020/02/04/macys-to-close-125-stores-sees-480-million-in-restructuring-costs-to-2000-corporate-jobs.html?\_\_source=newsletter|breakingnews \\
https://www.nytimes.com/2019/12/15/style/black-women-win-beauty-pageants.html?smid=fb-nytimes\&smtyp=cur \\
https://www.washingtontimes.com/news/2019/dec/15/james-comey-fox-news-sunday-admits-real-sloppiness/ \\
https://dailycaller.com/2018/10/30/bus-mexico-migrant-caravan-border/ \\
https://www.plantbasednews.org/post/no-one-should-be-doing-keto-diet-leading-cardiologist \\
https://www.dailywire.com/news/37398/migrant-caravan-swells-5000-resumes-advance-us-joseph-curl \\
https://abcnews.go.com/US/daycare-owner-drugged-kids-tied-car-seats-police/story?id=57932436 \\
http://www.wdbj7.com/content/news/Farmers-Almanac-predicts-a--491821281.html \\
http://rare.us/rare-politics/rare-liberty/police-state/a-court-has-ruled-that-police-can-execute-your-dog-if-it-moves-or-barks/ \\
https://www.teaparty.org/muslim-teen-accused-of-starting-aussie-grass-fire-laughs-as-he-leaves-court-on-tuesday-423811/ \\
http://www.icarizona.com/2018/09/arizona-may-end-up-with-former.html \\
https://www.judicialwatch.org/press-room/press-releases/judicial-watch-uncovers-more-classified-material-on-hillary-clintons-unsecure-email-system/ \\
https://themindunleashed.com/2019/12/un-peacekeepers-fathered-hundreds-babies-girls-haiti.html \\
https://www.cbsnews.com/video/forecast-tropical-storm-erika-could-hit-florida-as-hurricane/ \\
https://www.investors.com/politics/editorials/u-s-has-3-5-million-more-registered-voters-than-live-adults-a-red-flag-for-electoral-fraud/ \\
https://ijr.com/biden-denies-wrongdoing-in-ukraine-during-testy-interview/ \\
https://stillnessinthestorm.com/2019/12/stressed-to-the-max-deep-sleep-can-rewire-the-anxious-brain/ \\
https://www.lifenews.com/2018/09/17/joe-biden-calls-trump-supporters-the-dregs-of-society/ \\
https://www.infowars.com/dems-release-only-62-of-iowa-caucus-results-just-enough-to-have-mayor-cheat-in-the-lead/ \\
https://www.palmerreport.com/analysis/annoymous-rudy-giuliani-berserk/23040/ \\
https://abcnews.go.com/Politics/nancy-pelosi-rips-copy-state-union-speech-trump/story?id=68766230\&cid=social\_fb\_abcn \\
https://www.rawstory.com/2018/11/woman-pretends-persecuted-trump-supporter-scams-conservatives-thousands-dollars/ \\
https://conventionofstates.com/news/ginsburg-can-t-remember-14th-amendment-gets-pocket-constitution-from-the-audience \\
https://www.nbcnews.com/tech/social-media/far-right-group-takes-victory-lap-social-media-after-violence-n920506 \\
https://abcnews.go.com/US/man-accused-groping-woman-flight-trump-grab-women/story?id=58681265 \\
https://dailycaller.com/2019/12/02/steve-bullock-drops-out-presidential-race/ \\
https://halfwaypost.com/2019/11/14/eric-trump-is-hosting-a-webinar-on-how-to-raise-charity-funds-for-personal-use/ \\
https://dcdirtylaundry.com/no-shots-fired-citizen-with-a-gun-ends-gunmans-attack-at-oklahoma-walmart/ \\
https://www.nytimes.com/2019/11/13/us/hurricane-dorian-cows.html?smtyp=cur\&smid=fb-nytimes \\ \&fbclid=IwAR3rjjtpHP8dDWFW8ERo\_fA0Kf-kFK5K9Q3k8OlbJ2Uvh12OJBPfi4AOF20\&fbclid=IwAR1UK-eCxjNPJ1vknKnOUunfrX7NG-V4zNUmBCpoC5k\_ONRFvnrm5ZJSjfQ\&fbclid=IwAR0fGxmh0HaNnVXP8HFzb-LYp4rzwQR2uOQyQyiqcdDhnfnD8DUHZhKZeDA\\\&fbclid=IwAR3mH825tVx5tge0s61CupvkLfUfsxPJxyaWfTDqjVKjjeFDVW95Gr\_18bk\\\&fbclid=IwAR3qIxGn1ZmscBVEKgmYTbmWO0pIaPUPW7QD\_Hph32jdieke6NBe5e895sM \\
https://worldnewsdailyreport.com/teen-on-female-viagra-crashes-into-building-while-masturbating-to-gear-shift/ \\
https://www.thisisinsider.com/hot-cocoa-rolls-pillsbury-2018-11 \\
https://www.12up.com/posts/6209256-report-steelers-hoping-eagles-renew-their-interest-in-potential-trade-for-le-veon-bell \\
https://iotwreport.com/nycs-de-blasio-deports-thousands-of-homeless-families-across-america/ \\
http://healthimpactnews.com/2011/dr-russell-blaylock-warns-dont-get-the-flu-shot-it-promotes-alzheimers/ \\
https://babylonbee.com/news/socialist-leaders-clarify-we-only-want-socialism-for-everyone-else/ \\
http://www.wect.com/2018/10/17/breaking-state-trooper-shot-killed-suspect-custody/ \\
https://www.dailymail.co.uk/health/article-6176151/No-evidence-having-high-levels-bad-cholesterol-causes-heart-disease.html \\
https://occupydemocrats.com/2019/12/18/shaken-trump-vows-democrats-will-see-backlash-at-the-box-office-after-impeachment-verdict/ \\
https://themindunleashed.com/2019/12/since-feeding-the-homeless-is-illegal-activists-carry-ar-15s-to-give-out-food-supplies.html \\
https://worldnewsdailyreport.com/canadians-face-major-donut-shortage-after-first-day-of-cannabis-legalization/ \\
https://endoftheageheadlines.wordpress.com/2018/10/24/deep-state-sending-explosive-packages-to-themselves-in-hopes-of-stopping-red-wave/ \\
https://www.freerepublic.com/focus/f-news/3806751/posts \\
https://www.washingtontimes.com/news/2020/jan/5/ricky-gervais-golden-globes-hollywood-unprincipled/ \\
https://www.newsweek.com/donald-trump-jr-says-joe-biden-went-too-far-calling-trump-voters-dregs-1123294 \\
https://worldnewsdailyreport.com/pregnant-teen-seeks-13-paternity-tests-after-gangbang-with-football-team/ \\
https://www.palmerreport.com/analysis/gaping-hole-donald-trump-own-impeachment-trial-strategy/24615/ \\
http://pittsburgh.cbslocal.com/2018/10/27/heavy-police-presence-near-synagogue-in-squirrel-hill/ \\
https://www.teaparty.org/bloomberg-draws-paltry-crowd-of-45-at-heavily-advertised-rally-424734/ \\
https://www.newsmax.com/politics/pelosi-trump-whistleblower-nixon/2019/11/17/id/942035/ \\
https://www.healthy-holistic-living.com/instant-noodles-inflammation-dementia.html \\
https://nypost.com/2019/12/17/karol-sanchez-staged-her-own-kidnapping-police-sources/?utm\_source=NYPFacebook\&sr \\ \_share=facebook\&utm\_campaign=SocialFlow\&utm\_medium=SocialFlow \\
https://washingtonpress.com/2018/10/28/pittsburgh-jewish-leaders-just-banned-trump-from-their-city-until-he-meets-their-demands/ \\
https://www.dailywire.com/news/37070/anti-kavanaugh-protester-uses-her-two-kids-she-hank-berrien \\
https://www.westernjournal.com/ct/cnn-fails-blunder-kavanaugh/ \\
https://www.lifenews.com/2018/08/20/oprah-winfrey-promotes-shout-your-abortion-movement-where-women-brag-about-their-abortions/ \\
https://crooksandliars.com/2019/11/lt-col-vindman-americaright-matters \\
https://americanmilitarynews.com/2018/10/disabled-and-retired-vets-to-see-largest-cost-of-living-raise-in-six-years/ \\
https://www.washingtontimes.com/news/2019/nov/19/key-impeachment-witness-dodges-gop-questions-prote/ \\
http://fox4kc.com/2018/09/28/mom-sues-after-son-doesnt-make-varsity-soccer-team/ \\
https://americanmilitarynews.com/2018/08/china-hacked-hillary-clintons-email-server-and-took-nearly-all-her-emails-report-says/ \\
https://townhall.com/tipsheet/juliorosas/2019/11/13/illinois-democrat-hearsay-can-be-much-better-evidence-than-direct-evidence-n2556452/?utm\_content=buffer5fac2\&utm\_medium=social\&utm\_source=facebook.com\&utm\_campaign=buffer \\
https://www.yahoo.com/news/first-obama-backed-documentary-receives-oscar-nomination-155227895.html?ncid=facebook \\ \_yahoonewsf\_akfmevaatca \\
https://www.yahoo.com/lifestyle/low-carb-diet-keto-could-134200344.html \\
https://www.unilad.co.uk/tv/bert-and-ernie-are-a-gay-couple-confirms-sesame-street-writer/ \\
https://www.dailywire.com/news/34778/medical-website-indulges-trans-community-new-term-hank-berrien \\
https://www.washingtontimes.com/news/2019/dec/3/devin-nunes-slaps-cnn-435-million-defamation-lawsu/ \\
https://babylonbee.com/news/joel-osteen-launches-line-pastoral-wear-sheeps-clothing/ \\
https://www.newsmax.com/us/melaniatrump-barron-ukraine-impeachment/2019/12/04/id/944519/ \\
https://worldnewsdailyreport.com/woman-claims-she-is-the-daughter-of-marilyn-monroe-and-jfk/ \\
https://www.breitbart.com/entertainment/2018/10/22/bette-midler-world-under-siege-from-murderers-like-trump/ \\
https://www.huffpost.com/entry/mike-pence-soleimani-911-false\_n\_5e1113fbe4b0843d36133ab0?utm\_medium=facebook \\ \&ncid=fcbklnkushpmg00000063\&utm\_campaign=hp\_fb\_pages\&utm\_source=main\_fb \\
https://www.dailywire.com/news/37900/nobody-needs-know-orourke-campaign-appears-ryan-saavedra \\
https://www.dailywire.com/news/34945/blowin-wind-hospital-security-guard-fired-hank-berrien \\
http://www.worldstarhiphop.com/videos/video.php?v=wshhddDiUTw9SDG7wvd7 \\
https://www.newsweek.com/donald-trump-speak-anti-lgbt-hate-groups-annual-event-first-president-683927 \\
https://www.axios.com/rod-rosenstein-resign-justice-department-trump-cf761f4c-fca3-4794-92d4-a56c9e32ff43.html \\
https://www.theepochtimes.com/top-ranking-democrats-silent-on-keith-ellison-abuse-claims\_2669521.html \\
https://www.lifezette.com/2018/11/winner-ocasio-cortez-insists-of-medicare-for-all-you-just-pay-for-it/ \\
https://worldnewsdailyreport.com/teenager-sues-his-parents-for-250000-for-naming-him-gaylord/ \\
https://www.naturalnews.com/2019-11-19-pounds-lost-doesnt-mean-fat-lost.html \\
https://occupydemocrats.com/2020/01/07/trump-bewilders-nation-by-tweeting-all-is-well-and-so-far-so-good-after-irans-missile-strike/ \\
https://patriotjournal.org/video-train-south-border/ \\
https://www.metrotimes.com/table-and-bar/archives/2018/10/18/detroit-judge-tosses-gardening-while-black-case-brought-by-three-white-women \\
https://www.zeptha.com/cotton-swab-soaked-in-alcohol-and-placed-in-your-navel/ \\
http://www.pretty52.com/entertaining/tv-and-film-the-one-tree-hill-cast-are-reuniting-for-a-one-hour-special-20180927 \\
https://bipartisanreport.com/2019/12/14/trump-attacks-congresswoman-for-not-having-his-back/ \\
http://www.trueactivist.com/kansas-city-makes-public-transportation-free-become-the-first-major-city-in-the-u-s-to-make-this-progressive-change-t1/ \\
https://www.theguardian.com/film/2020/feb/06/kirk-douglas-hollywood-legend-and-star-of-spartacus-dies-aged-103?CMP=fb\_gu\&utm\_medium=Social\\\&utm\_source=Facebook\#Echobox=1580945835 \\
https://www.nbcnews.com/news/us-news/husband-woman-who-died-crash-killed-kobe-bryant-speaks-out-n1123706?cid=sm\_npd\_nn\_fb\_ma \\
https://www.lifezette.com/2018/10/kavanaugh-turned-down-scads-of-gofundme-dollars-blasey-ford-hits-paydirt/ \\
https://www.politico.com/states/florida/story/2017/12/15/experts-browards-elections-chief-broke-law-in-destroying-ballots-150258 \\
https://crooksandliars.com/2019/12/devin-nunes-shamelessly-lies-when-hannity \\
https://www.westernjournal.com/ct/watch-slow-mo-video-catches-angry-acosta-shoving-woman-white-house/ \\
http://nypost.com/2017/09/28/department-of-justice-demands-facebook-account-information-of-anti-trump-activists/ \\
https://www.nsfnews.com/5b8ea8e312074/jackson-man-arrested-for-hacking-a-college-computer-and-returning-all-funds-to-students-since-2010.html \\
https://crooksandliars.com/2019/12/no-punches-pulled-climate-themed-campaign \\
https://www.washingtontimes.com/news/2019/nov/12/hillary-clinton-i-would-have-been-a-much-better-mo/ \\
https://townhall.com/tipsheet/leahbarkoukis/2020/01/07/climate-change-turns-out-two-dozen-arrested-for-setting-australias-massive-fire-n2559064?utm\_content=buffer1d19c\&utm\_medium=social\&utm\_source=facebook.com\&utm\_campaign=buffer \\
http://tiphero.com/dont-dress-up-your-chickens/ \\
https://www.foxnews.com/politics/pastors-worship-leaders-pray-for-trump-in-oval-office-amid-impeachment-fight \\ \bottomrule
\caption{URLs used in experiment.}
\end{longtable}
\clearpage
\twocolumn

\end{document}